\documentclass[preprint,12pt]{elsarticle}
\usepackage{natbib}
\setcitestyle{numbers,sort&compress,square}
\usepackage{amssymb}

\usepackage{float, graphicx}
\usepackage{siunitx}
\DeclareSIUnit\Molar{M}
\usepackage{etoolbox}
\usepackage{lmodern}
\usepackage{amsmath}
\usepackage{ragged2e}
\usepackage{upgreek}
\usepackage{blindtext}
\usepackage{booktabs}
\usepackage{subcaption}
\usepackage{multicol,caption}
\usepackage{nccmath, mathtools}
\usepackage{braket} 
\usepackage{textcomp}
\renewcommand\harvardurl[1]{\textbf{URL:} \url{#1}}
\usepackage{hyperref}

\journal{Radiation Physics and Chemistry}
\begin{document}
\begin{frontmatter}

\title{Monte Carlo Simulation and Dosimetric Analysis of Gold Nanoparticles (AuNPs) in Breast Tissue}

\author[inst1,inst2]{Pedro Teles}\corref{cor1}\corref{cor2}\corref{cor3}\corref{cor4}
\author[inst1]{Catarina Dias}
\author[inst1,inst3]{João H. Belo}
\author[inst1,inst3]{Célia Sousa}
\author[inst4]{Paula Boaventura}
\author[inst2]{Isabel Bravo}
\author[inst2]{João Santos}
\author[inst5]{Marissa Rylander}

\affiliation[inst1]{organization={DFA/FCUP, Porto, Portugal}}
\affiliation[inst2]{organization={IPO-Porto, Portugal}}
\affiliation[inst3]{organization={IFIMPUP, Porto, Portugal}}
\affiliation[inst4]{organization={i3S, Porto, Portugal}}
\affiliation[inst5]{organization={Department of Mechanical Engineering at the University of Texas, Austin}}
\cortext[cor1]{Corresponding author: ppteles@fc.up.pt}
\cortext[cor2]{The list of authors is different from the version submitted to RPC; the request for an author list change was made with the agreement of all authors.}
\cortext[cor3]{The errors found and sent as an erratum  to the editorial office have also been corrected in this version. These errors did not affect the conclusions.}
\cortext[cor3]{The paper was not accepted in its current form. The authors will try to address the comments made by the reviewers and provide an updated script. Some of their concerns are already taken into account in this current iteration.}
\begin{abstract}
\sloppy     Precise radiation delivery is critical for effective radiotherapy, and gold nanoparticles (AuNPs) have emerged as promising tools to enhance local dose deposition while sparing the surrounding healthy tissue.
    In this study, the PENELOPE Monte Carlo code was used to investigate the dosimetry of AuNPs under different conditions and models.
    The Dose Enhancement Ratio (DER) was studied in water and breast tissue with spherical shapes and in agreement with previously published results. To further analyse the physical interactions of the particles around the AuNP, a Phase Space File (PSF) in a volume around the AuNPs was created. This showed that larger AuNPs lead to increased doses, as expected, yielding DER values exceeding 100 times. Finally, results reveal that in the volume surrounding the AuNP, 80\% of emitted electrons originate from photoelectric absorption, leading to  Auger electron emission cascades which were analysed in detail. It was also possible to establish a direct relation between number of secondaries and the particle volumes.
    The Local Effect Model (LEM) was  used to determine survival curves in AuNPs of different sizes at different gold concentrations. The last part of this work consisted in analysing a distribution of AuNPs within a flattened cell typical of clonogenic assays where a log-normal distribution of dose was observed. This led to the development of a new, mechanistic, Local Effect Model which, if further validated, can have further applications in-vitro and in-silico. 

\end{abstract}
   
     
\end{frontmatter}

\section{Introduction}

Cancer is a global problem demanding continuous developments in both treatment and diagnostic techniques. In the fight against cancer, a notable advance is that at least in Europe and in the US, the standardized mortality rate has slowly but steadily declined recently  \cite{num1,num2,num3,num4}. This decline can be explained by the ongoing technological developments in cancer therapies and early diagnostics \cite{num5}. The “War on Cancer” has not yet been won, but innovative treatments can lower mortality rates and improve the patients’ quality of life during and after treatment. Innovative treatments focus on targeting tumours while sparing the surrounding healthy tissue, localizing efficacy, and reducing side effects.

Gold nanoparticles (AuNPs) are a promising solution, with various therapeutic applications, including Gold Nanoparticle Assisted Radiation Therapy (GNRT) \cite{num10}, and Photo Thermal Therapy (PTT) \cite{num11}. The typical X-ray energies needed for irradiation (20-150 keV)  show theranostic potential,  which has already shown promise in preclinical animal studies \cite{num13}. 

In the case of GNRT, the rationale is that Gold (Au) has a high photoelectric absorption coefficient at diagnostic X-ray energies, as well as elevated Auger and Coster–Kronig (C-K) electron emission yields. As such, AuNPs release a cascade of such electrons, leading to a localized dose enhancement, often reaching several orders of magnitude, within a few hundreds of nanometres surrounding the particle. Dose enhancement ratios are commonly assessed using Monte Carlo simulations \cite{num12,corrigendum}.

AuNPs come in different sizes and shapes which can be further modified through surface coatings and functionalizations \cite{num7}. Furthermore, there are various biochemical entities that they can be conjugated with \cite{num6} with high biocompatibility, displaying controllable patterns \cite{num8}. AuNPs can target specific tumour cells and unleash their therapeutic potential upon irradiation \cite{num8, num9}.

Gold has a high photoelectric cross-section, and AuNPs exhibit heightened yields of Auger and Coster-Kronig electrons \cite{nist_cs}. These electrons have nanometric ranges, depositing energy in their vicinity and amplifying biological damage in surrounding tissues \cite{num21}.
In fact, for gold, for energies below the ”K-edge”,   Auger electron emission will dominate. Their emission probability increases gradually from the inner to the outer shells:  4.1\% (K), 74.6\% (L), up to 98.5\% (M) \cite{review_2018}. Finally, the high emission of Auger electrons leads to an accumulation of positive charge around the nanoparticle, increasing its chemical reactivity with the surrounding medium. This, in turn, attracts electrons from nearby water molecules and biomolecules, resulting in rapid atomic relaxation processes and the emission of extra electrons \cite{review_2018}.

The Monte Carlo method is the primary technique for dosimetric calculations, providing insights into secondary electron processes and dose enhancement due to AuNPs \cite{sug_joao}. 

Several recent studies have focused on benchmarking dose enhancement ratios (DER) using different codes \cite{num12, corrigendum}, 

Nanoparticle location and concentration within the cell are critical factors in cell radiosensitivity. The precise location of AuNPs is essential because Auger electrons have limited ranges \cite{MC_vitro}. The non-uniform dose distribution induced by AuNPs was a crucial factor in several studies, motivating the implementation of the Local Effect Model (LEM) \cite{Brown2017, LEM_review,LEM_protons}. The LEM has shown promising results, with strong agreement between its predicted curves and experimental irradiation data. This method has been the subject of recent debate in which its validity has been discussed \cite{Rabus1,Rabus2}. 

This work is an attempt to provide further and more detailed information about the physical interactions behind the dose enhancement provoked by the AuNPs, and their potential application in cancer treatment, potentially breast cancer \cite{zhang2012size}. 

For this purpose the PENELOPE model was first validated against benchmarked results in the literature \cite{num12, corrigendum}. After validation, we shift to breast tissue, and recalculate the DERs.

Shifting to a 50 kVp source, we  incorporate a spherical shell Phase Space File (PSF) analysis to systematically track particle interactions, providing a rigorous assessment of dose enhancement effects and the physical processes behind them. 

After, the local effect model (LEM) was used to determine  survival curves and sensitization enhancement, integrating spatially resolved dose distributions with a previously validated model \cite{Brown2017}, attempting to shed light on the use of AuNPs coupled with low energy x-ray sources in cell sensitization.

Considering the need for in-vitro clonogenic-assay like dose benchmarks, the final step of this study involved the use of a flattened cell model in which a pattern of AuNPs was placed, irradiated with a beam with the same 50 kVp spectrum \cite{cell_au}, after which the dose deposition distribution was analyzed.

This allowed the development a new framework for a modified Local Effect Model that is further discussed in the text.

\section{Methodology}
    \subsection{PENELOPE}

The Monte Carlo code used for simulating electron-photon transport in this study was the 2018 version of PENELOPE "PENetration and Energy LOss of Positrons and Electrons" \cite{penelope}, making use of the PenEasy framework \cite{penEasy}

\sloppy In the input file,  a table for transport parameters, including material, absorbed energy ($\mathrm{E_{ABS}}$), parameters for hard elastic events ($\mathrm{C_1}$ and $\mathrm{C_2}$), cut-off energies for hard inelastic collisions (W$_{CC}$) and for hard bremsstrahlung (W$_{CR}$), and finally, the maximum permissible step length for electrons (DS$_{1unit{MAX}}$) must be provided. The values used for these parameters are provided in table \ref{tab:penelope_params} of the Supplementary Material.

It is noteworthy that PENELOPE is a Monte Carlo code that employs a hybrid method between analog and condensed history simulations. The values chosen were a compromise between keeping the simulation analog and computer time.

The code was run on a typical home laptop. $10^7$ particle histories were generated to ensure good statistics ($2\sigma$ below 2\%).

\subsection{Simulation methodology}
\subsubsection{Dose enhancement ratio in water}

The first step in this work was to validate the model used in the simulations. For this, similar specifications to a previous study were used \cite{num12,corrigendum} to evaluate the dose enhancement of a single AuNP surrounded by water upon irradiation with an X-ray beam, which allowed for a comparison and validation of the results. SpekPy \cite{spekpy} was used to produce two X-ray spectra in the diagnostic range - 50 kVp and 100 kVp. An anode angle of $20^\circ$ and a filtration setup of 0.8 mm of beryllium (Be) + 3.9 mm of aluminum (Al) was used. A figure of the spectra used in given in Figures \ref{fig:spectrum50} and \ref{fig:spectrum100} in the Supplementary Material.

The AuNP was set at the origin and the "source box" was defined with its centre at coordinates (0, 0, -100) \SI{}{\micro\metre}, consisting of a square with a side length 10 bmlarger than the AuNP, emitting a parallel beam of photons following the 50 kVp or the 100 kVp spectrum. 

Given that for the validation we used AuNPs of 50 and 100 nm diameters,  the source side length was of 60 nm and 110 nm  respectively. Concentric 25 nm width subshells were used to score the dose as a function of distance to the AuNP. The results were then compared with those by Li \textit{et al} \cite{num12,corrigendum} .

\subsubsection{DER in breast medium}

After the validation of the model, water was replaced with breast tissue. The specifications of breast tissue were taken from the ICRU Report 44 (1989) \cite{mat_breast}. The simulated AuNP diameters were 12.5, 25, 50, and 100 nm, with the same square-type source  (50 kVp X-ray spectra)  10 nm larger than the diameter, at 22.5, 35, 60, and 100 nm, respectively.

\subsubsection{Phase Space File generation}

 Following the methodology of \cite{LEM_protons}, we decided to implement a similar method but leveraging the spherical geometry of the AuNP, tallying the detailed interaction structure around a spherical shell, to address potential biased underscoring when using a linear approach.

\begin{figure}[h!]
    \centering
    \includegraphics[width=0.6\linewidth]{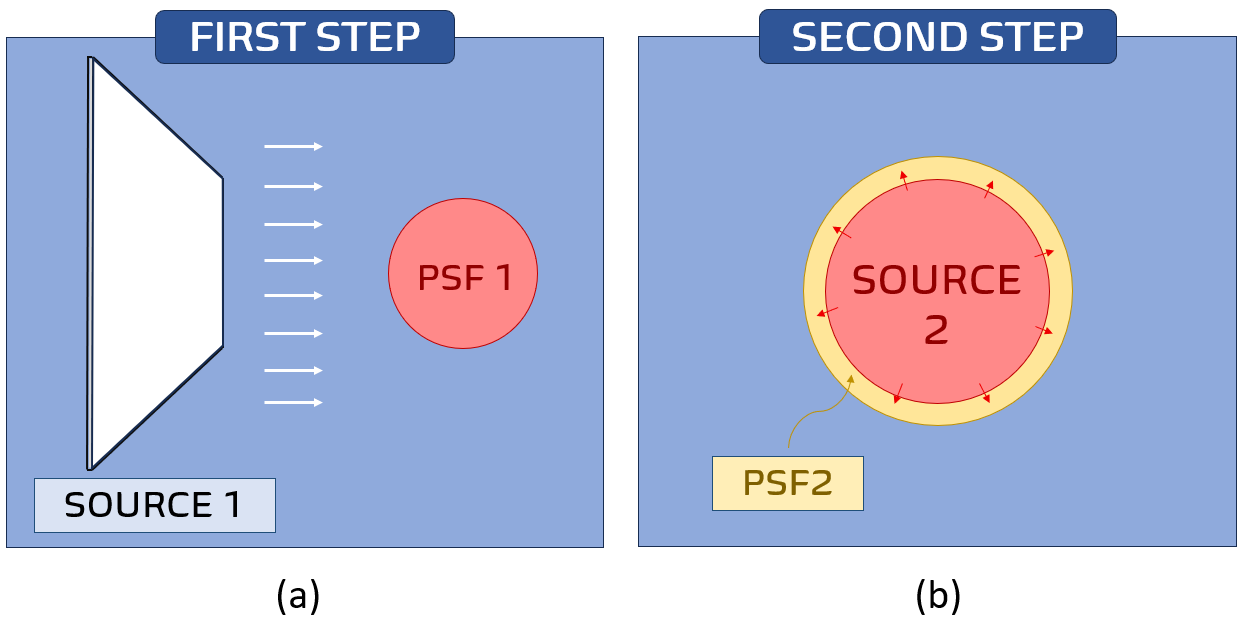} \captionof{figure}{Irradiation scheme method for the creation of the PSFs. Elements in this scheme are not to scale.}
    \label{fig:psf_steps}
 \end{figure}

To do this, two PSFs were created. One to capture all the information of the particles entering the AuNP volume (or alternatively the baseline - breast medium) after irradiation with the source, as illustrated in Figure \ref{fig:psf_steps}(a), and a second one (made of breast tissue), in which the AuNP (or breast tissue) becomes itself the source, and the surrounding  diameter, larger by 20\%  (which we named “surrounding halo”) scores the outcoming particles in a second  PSF. The information of the particles leaving the AuNP was then analysed in detail.

An illustration of the irradiation conditions for both PSFs is depicted in figure \ref{fig:psf_steps}. To note that the figure is not to scale.

This split into two stages was made as each PSF's absorption energies need to be set at "infinity" to avoid duplication of secondary particles in the detection material (ie the material where the PSF file will be built), as recommended in the PenEasy manual \cite{penelope,penEasy}.

The observed secondary particle creation is given in detail in tables \ref{tab:tissue_secondary_particles} and \ref{tab:secondary_particles} of Supplementary Material.

\subsubsection{Survival Curves}

An estimation of survival curves in the presence of AuNPs was made by making use of
the Local-Effect-Model (LEM) formalism \cite{Brown2017}. 

The LEM considers the enhancement provoked by the AuNP:

\begin{equation}
     S_{enhanced} = \exp\bigg[-\left(\alpha (D_{enhanced})+\beta (D_{enhanced})^2\right)\bigg], 
     \label{eq:S_enhanced}
\end{equation}

where $\alpha$ and $\beta$ are the typical parameters of the linear-quadratic model, and $D_{enhanced}$ is the enhanced dose, determined as:

\begin{equation}
      D_{enhanced} = \sum_i\left[\left(D_i+\Delta N \cdot NP \cdot   d'_i\right)\cdot \frac{dV_i}{V_t} \right], 
     \label{eq:S_enhanced}
\end{equation}

 $\Delta N$ is the number of  ionisations, $NP$ is the number of nanoparticles, and $d'_i$ is the dose per source particle $d_i$,  in each shell "i", divided by the probability of an "extra" ionization provoked in the nanoparticle per source particle (i.e., the simulation itself tallies all ionisations, but only the actual ionisations provoked by the nanoparticle need to be accounted for).

Then we get:
 
{\scriptsize
\begin{equation}      
        S_{enhanced} = \exp\bigg[-\left(\alpha D+\beta D^2\right) - \Delta N \cdot NP \sum_i \left(   \big[\left( \alpha  d'_i + 2\beta D  d'_i \right)\big] \cdot \frac{dV_i}{V_t} +   \beta \left(d'_i\cdot \frac{dV_i}{V_t}\right)^2\right) \bigg].
     \label{eq:S_enhanced_III}
\end{equation}
}

The term $2\beta D$ can be discarded as the
probability of a baseline and enhancement track hitting the same DNA
site is negligible at  the considered  photon energies \cite{Brown2017}. Therefore, this equation becomes:

{\footnotesize
\begin{equation}
        S_{enhanced} = \exp\bigg[-\left(\alpha D+\beta D^2\right) - \Delta N \cdot NP \sum_i \left(    \alpha  d'_i  \cdot \frac{dV_i}{V_t} +  \cdot  \beta \left(d'_i\cdot \frac{dV_i}{V_t}\right)^2\right) \bigg].
     \label{eq:S_enhanced_IV}
\end{equation}
}

To determine the number of ionisations,  CPE is assumed to establish a relation between dose and energy fluence of the photons. With this, we can determine the number of particles necessary to deposit 1 Gy in the tissue volume to normalize the raw data coming from the PENELOPE file (in eV/g/source particle).

\begin{equation}
\begin{aligned}
    K_c &= D \\
    &\Leftrightarrow \langle\left( \frac{\mu_{en}}{\rho}_{tissue} \right)\rangle \psi = D \\
    &\Leftrightarrow \langle\left( \frac{\mu_{en}}{\rho}_{tissue}\right)E\rangle \phi  = D \\
    &\Leftrightarrow \langle\left( \frac{\mu_{en}}{\rho}_{tissue} \right)E\rangle \frac{N_{1Gy}}{A_{souce}} = D \\
    &\Leftrightarrow N_{1Gy} = \frac{A_{source} D}{<\bigg( \frac{\mu_{en}}{\rho}_{tissue} \bigg) E>} 
\end{aligned}
\label{eq:N_1Gy}
\end{equation}

In this equation, the term in parenthesis accounts for the mass energy absorption coefficient  $\langle\left( \frac{\mu_{en}}{\rho}_{\unit{tissue}}\right)E\rangle$ in tissue, weighted by the energy bins of the 50 keV spectrum, the term  $A_{Source}$ is the cross-sectional area of the source, and $N_{1Gy}$ represents the number of source particles necessary to deposit 1 Gy in the AuNP (which will be used for normalisation purposes).

The number of ionisations corresponds to the number of photons that interact with the NP (considering that each interaction will lead to an ionization), which are the ones not absorbed by it,

\begin{equation}
\Delta N = N_{1Gy} \cdot \frac{Area_{AuNP}}{Area_{source}}\cdot (1-e^{-\mu_{phot} \cdot  \bar{x}}),  
\label{eq:delta_N}
\end{equation}

where $\bar{x}$ represents  the mean chord length of the AuNP. 
This expression can be well approximated by a Taylor expansion up to the first order, as $\bar{x}$ is extremely small:

\begin{equation}
    \Delta N = N_{1Gy} \cdot \frac{Area_{AuNP}}{Area_{source}}\cdot \mu_{phot} \cdot  \bar{x}.  , 
    \label{eq:delta_N_II}
\end{equation}

Rewriting the term $N_{1Gy}$ explicitly:

\begin{equation}
    \Delta N = \frac{D_{1Gy}}{<\bigg( \frac{\mu_{en}}{\rho}_{tissue} \bigg) E>} \cdot Area_{AuNP} \cdot \mu_{phot} \cdot  \bar{x}.
    \label{eq:delta_N_III}
\end{equation}

Noting that $\bar{x}=\frac{Vol_{AuNP}}{Area_{AuNP}}$, we can write:

\begin{equation}
    \Delta N = \frac{D_{1Gy}}{<\bigg( \frac{\mu_{en}}{\rho}_{tissue} \bigg) E>}  \cdot \mu_{phot} \cdot Vol_{AuNP},
    \label{eq:delta_N_IV}
\end{equation}
meaning that the number of ionisations is proportional to the volume of the nanoparticle. This has clear implications, as it points to the shape of the NP being irrelevant as long as the volume is the same, while at the same time telling us that the number of ionisations is proportional to the volume of the AuNP \cite{rabus2021intercomparison}. 

Finally, equation \ref{eq:delta_N_II} can  be rewritten as:

\begin{equation}
    \Delta N = \Phi_{1Gy} \cdot Area_{AuNP}\cdot \mu_{phot} \cdot  \bar{x}, 
    \label{eq:delta_N_V}
\end{equation}

Where $\Phi_{1Gy}$ is the source particle fluence required to deposit 1 Gy at the AuNP site. The obtained value for the fluence  was  $\Phi_{1Gy} \approx 1.60 \times 10^{12}$ particles $\mathrm{cm^{-2}}$, and $N_{1Gy} \approx 58 $  particles.

The source was the same used for the validation of the 50 nm diameter AuNPs (a square source with a side length of 60 nm).

The concentrations used were fixed for all AuNP sizes at 0.1, 10.0 and 100.0 mM corresponding to 590, 5900, and 590000 NP for the 12 nm; 65, 652, and 65250 for the 25 nm ; 8, 82, and 8156 for the 50 nm, and finally 1, 10, and 1019 for the 100 nm.

To apply the LEM, the PENELOPE dose per shell was 
 summed per volume weight  over all concentric shells, capped at $r_{\text{low}}\le 5\;\upmu\text{m}$—approximately the radius of a breast-cell nucleus.  The corresponding reference volume is therefore \(V_{t}= \tfrac{4}{3}\pi(5\,\upmu\text{m})^{3}
       = 5.24\times10^{-10}\,\text{cm}^{3}\).

PENELOPE tallies provide $d'_i$ and $D_i$ in eV/g per primary photon.  To obtain survival curves as a function of macroscopic, clinically prescribed dose in Gy, we normalize these values using equation \ref{eq:delta_N_IV}, and by using the  spectrum-weighted mass absorption coefficient of breast tissue 
$\langle(\mu_{\mathrm{en}}/\rho)_{\mathrm{Au}}\rangle=1.260\times 10^{-1}\,$cm$^2$g$^{-1}$
and $(\mu/\rho)_{\mathrm{phot}}^{\mathrm{Au}}=19.75\,$cm$^2$g$^{-1}$ \cite{nist_cs} for ($\langle E\rangle=31\,$keV), and
$\rho_{\mathrm{Au}}=19.32\,$g cm$^{-3}$, we obtain
$\Delta N \approx 0.04 $ ionisations NP$^{-1}$ Gy$^{-1}$ for
50 nm particles, scaling with volume for the other diameters. This value is in line with other published results\cite{consequences, Yan2021} : Here we assume that:

\begin{equation}
    <\bigg( \frac{\mu_{en}}{\rho}_{tissue} \bigg) E> \approx <\bigg( \frac{\mu_{en}}{\rho}_{tissue} \bigg)>\cdot<E>,
\end{equation}

which holds true for these energy ranges.

Finally, we sweep over dose values between $0.1\le D_{\text{ref}}\le10$ Gy to obtain clinically relevant doses.

Each PENELOPE spherical histogram was binned up to $\Delta d=1\times10^{-4}$ Gy.  For every bin, the LQ expression in
Eq.~(\ref{eq:S_enhanced_IV}) was applied, and the results summed, giving
one survival value per prescribed dose.  The full set of points was
fitted with a LQ function to extract the effective parameters
$\alpha'$ and $\beta'$.

\subsubsection{Distribution of AuNPs in a cell}

\begin{figure}[h!]
    \centering
    \includegraphics[width=0.7\linewidth]{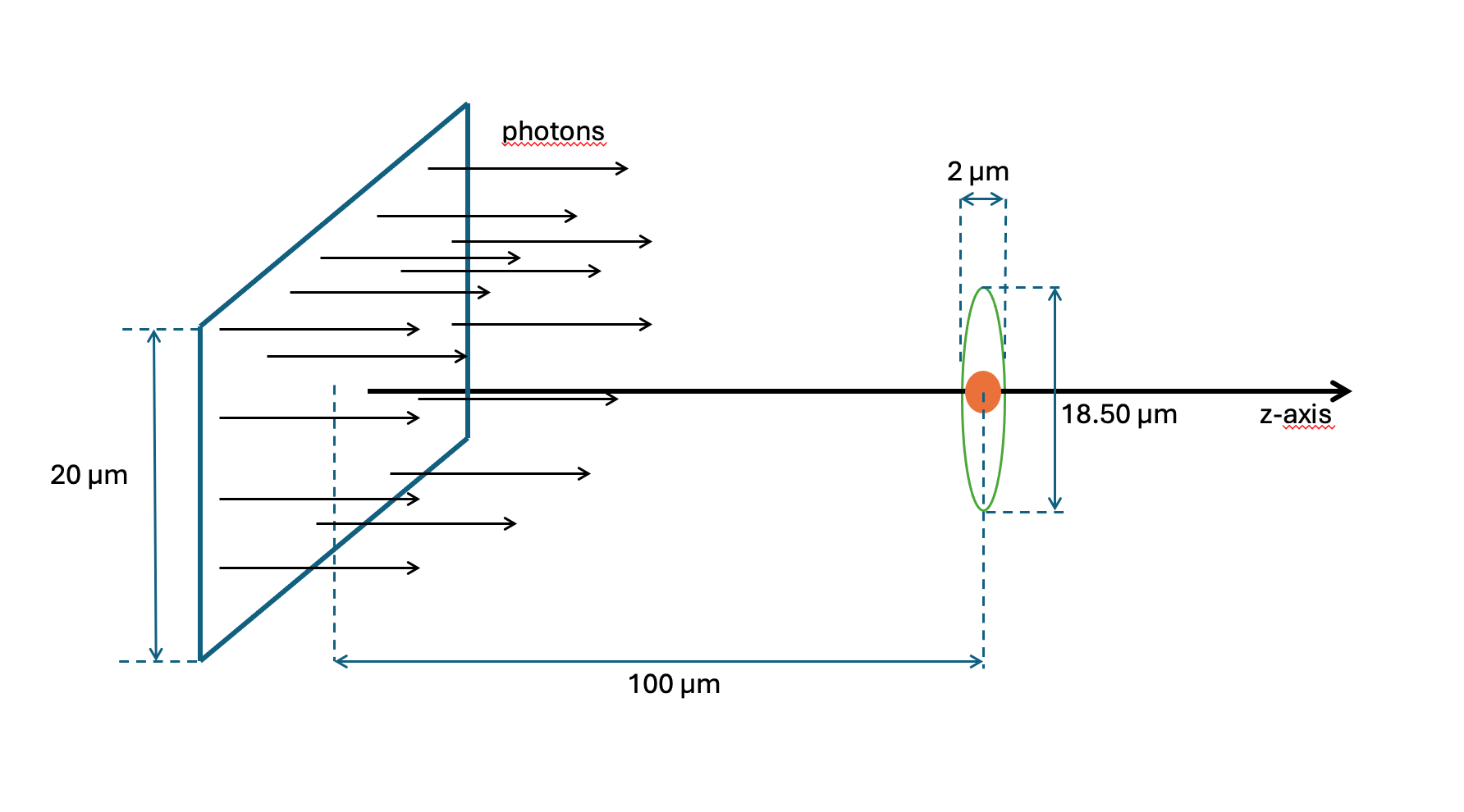} \captionof{figure}{Schematic representation of the irradiation geometry. The dimensions are not to scale.}
    \label{fig:irr_scheme}
 \end{figure}

The subsequent stage of this study involved assessing the impact of a distribution of AuNPs within a flattened cell in a typical in-vitro irradiation scenario \cite{cell_au}. Reproducing that setting, an elliptical cell geometry  was considered, with radii $(a, b, c)=(9.25, 4.25, 1) \ \unit{\micro m}$. Instead of a sphere, the nucleus was also described as an ellipsoid with $(a_n,b_n,c_n)=(4,4,0.1) \ \unit{\micro m}$. The source was placed at a distance of  $ 100 \ \unit{\micro m} $ from the cell's centre, forming a square with $20 \ \unit{\micro m}$ side in the $xy$-plane and emitting particles parallel to the $z$-axis, as shown in Figure \ref{fig:irr_scheme}. The previous 50 kVp X-ray spectrum was used.

Moreover, the "Tally spatial dose distribution" was used to score the dose, dividing the cell's nucleus into 200 voxels to compute the dose in each. In both $x$ and $y$ directions, there were 10 bins, while in the $z$ direction, there were two bins. A customized Python code analyzed the output data.
 
The absorbed dose in the voxels was determined for two different scenarios -  without gold and with 2000 AuNPs distributed along the inner surface of the cell. The coordinates of the AuNPs location were determined using the Python package "random.uniform". Because AuNPs are only expected to be scattered in the cytoplasm, a function named ”nucleus” was built, which returns 1 if the sample is inside the nucleus and 0 otherwise. The first initial approach was to use  2000 nanoparticles with a diameter of 50 nm and 100 nm, corresponding to a concentration of $8.15 \unit{mM}$ and $65.2 \unit{mM}$, respectively (see Appendix C of Supplementary Material). 

The dose obtained for each voxel of the flattened cell was then placed in a histogram to create a representative dose probability distribution reflecting the data.

\section{Results}

\subsection{Validation in water medium}

\begin{figure}[h!]
    \centering
    \begin{subfigure}{0.45\textwidth}
        \centering
        \includegraphics[width=\textwidth]{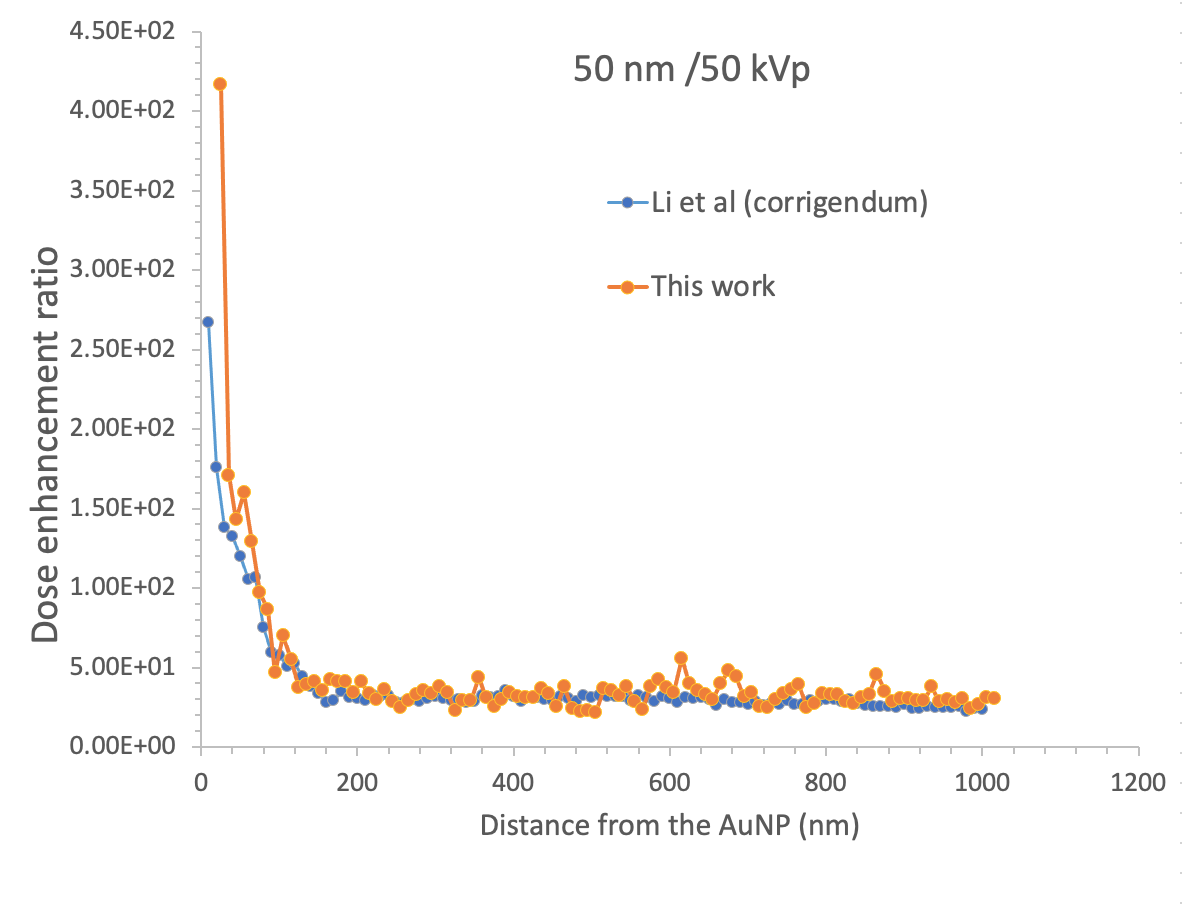}
        \caption{Comparison between our results and Li \textit{et al} \cite{corrigendum} for 50 nm/50 kVp }
        \label{fig:sub1}
    \end{subfigure}
    \hfill
    \begin{subfigure}{0.45\textwidth}
        \centering
        \includegraphics[width=\textwidth]{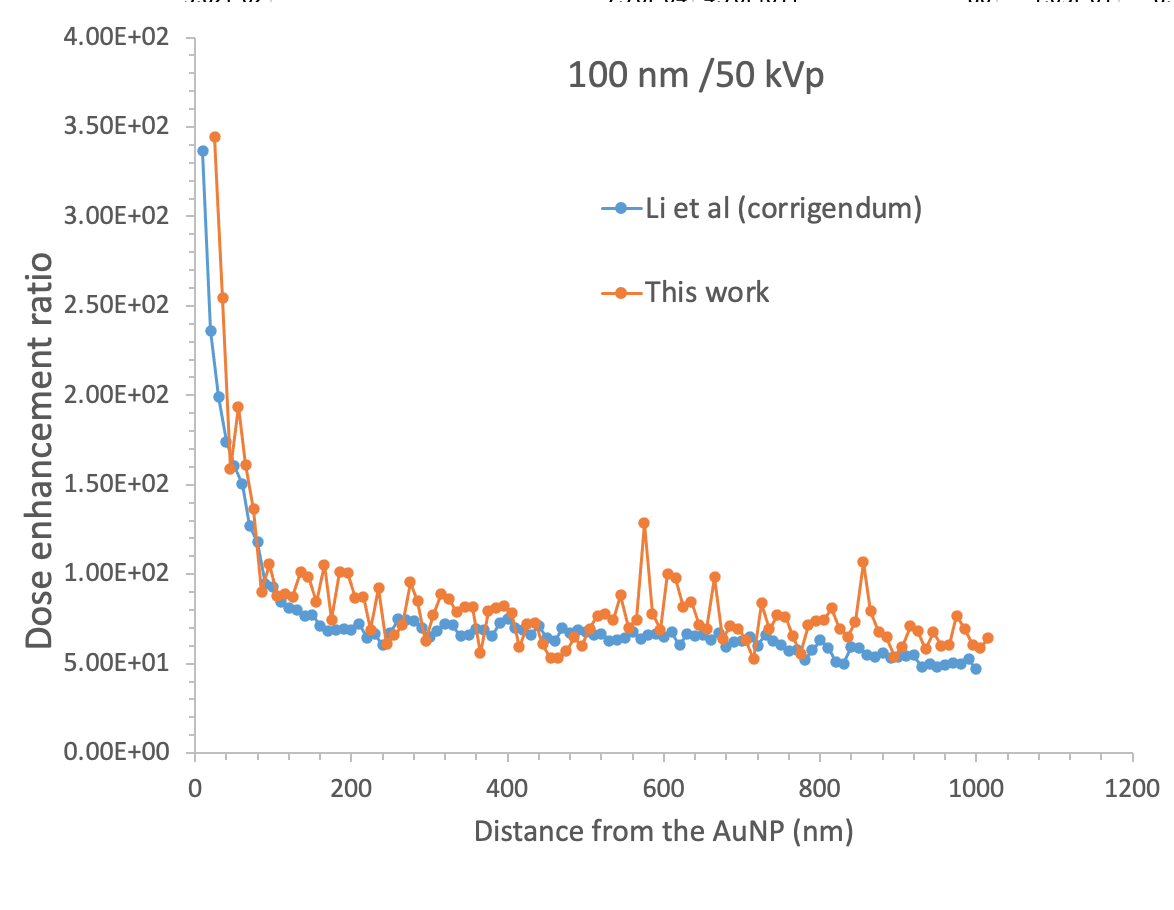}
        \caption{Comparison between our results and Li \textit{et al } \cite{corrigendum} for 100 nm/50 kVp}
        \label{fig:sub2}
    \end{subfigure}
    \caption{Dose-enhancement ratio (DER) versus radial distance from the nanoparticle surface.  Orange symbols: present work.  Blue symbols: Li \textit{et al.} \cite{corrigendum}.  Our first point is in the 0–25 nm bin ; Li \textit{et al.} plot in  0–10 nm bins.}
    \label{fig:both}
\end{figure}

The DER curves obtained in this study were benchmarked against the PENELOPE results of Li \textit{et al.} \cite{corrigendum}.  Figure~\ref{fig:both} overlays the two data sets for 50 nm and 100 nm AuNPs at 50 kVp.

Other DER vs distance curves were made for the different irradiation setups in combinations: 50 nm vs 50 kVp; and 100 kVp and 100 nm vs 50 kVp and 100 kVp, all yielding similar results, with results within less than 20\,\% from those of Li \textit{et al}.  

The higher statistical scatter in our data is expected, because of the use of 25 nm wide spherical shells versus 10 nm wide in Li \textit{et al}'s results. This is particularly relevant in the first point obtained for the 50nm/ 50 kVp result, which can be explained by the difference in binning. Apart from this first-point offset, the radial behaviour and long-range plateaux agree well with the comparable input parameter results of the same work \cite{artigo_2020,corrigendum}.

It should be pointed out that there are some differences in the irradiation set-up: in the case of this study a 50 kVp tungsten-anode spectrum generated with SpekPy (0.8 mm Be + 3.9 mm Al filtration) delivered as a 60 nm × 60 nm square micro-beam was used, whereas Li \textit{et al} employed the reference 50 kVp diagnostic spectrum and a circular pencil beam 60 nm in diameter. These small spectral and field-shape discrepancies could introduce minor deviations between the two data sets.

\subsection{Breast medium}

\subsubsection{Dose enhancement ratio}

\begin{figure*} [h!]
\centering
\begin{minipage}[t]{0.3\textwidth}
  \includegraphics[width=1.\linewidth, height=1.3\linewidth]{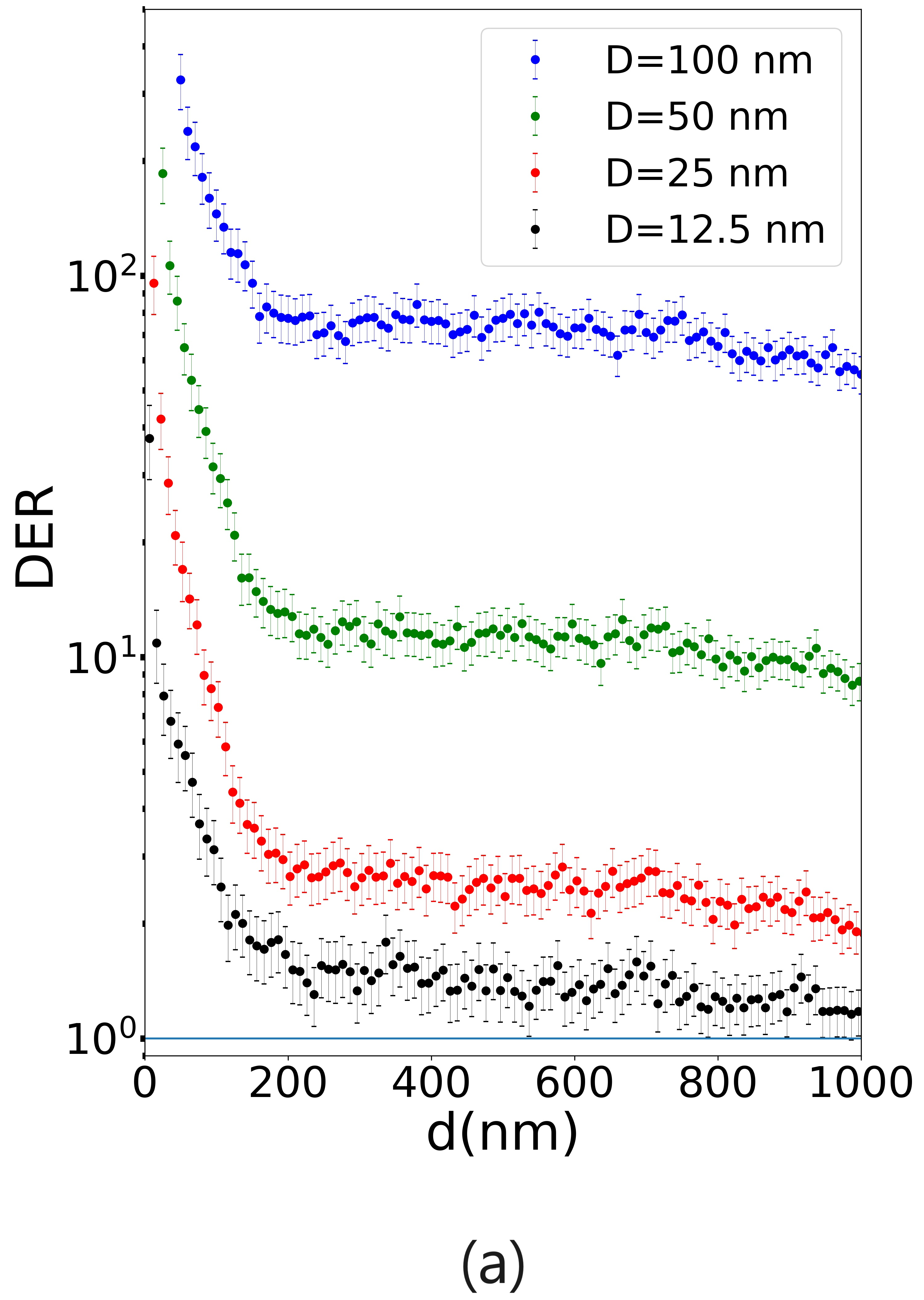}
\end{minipage}%
\hfill 
\begin{minipage}[t]{0.3\textwidth}
  \includegraphics[width=1.\linewidth, height=1.3\linewidth]{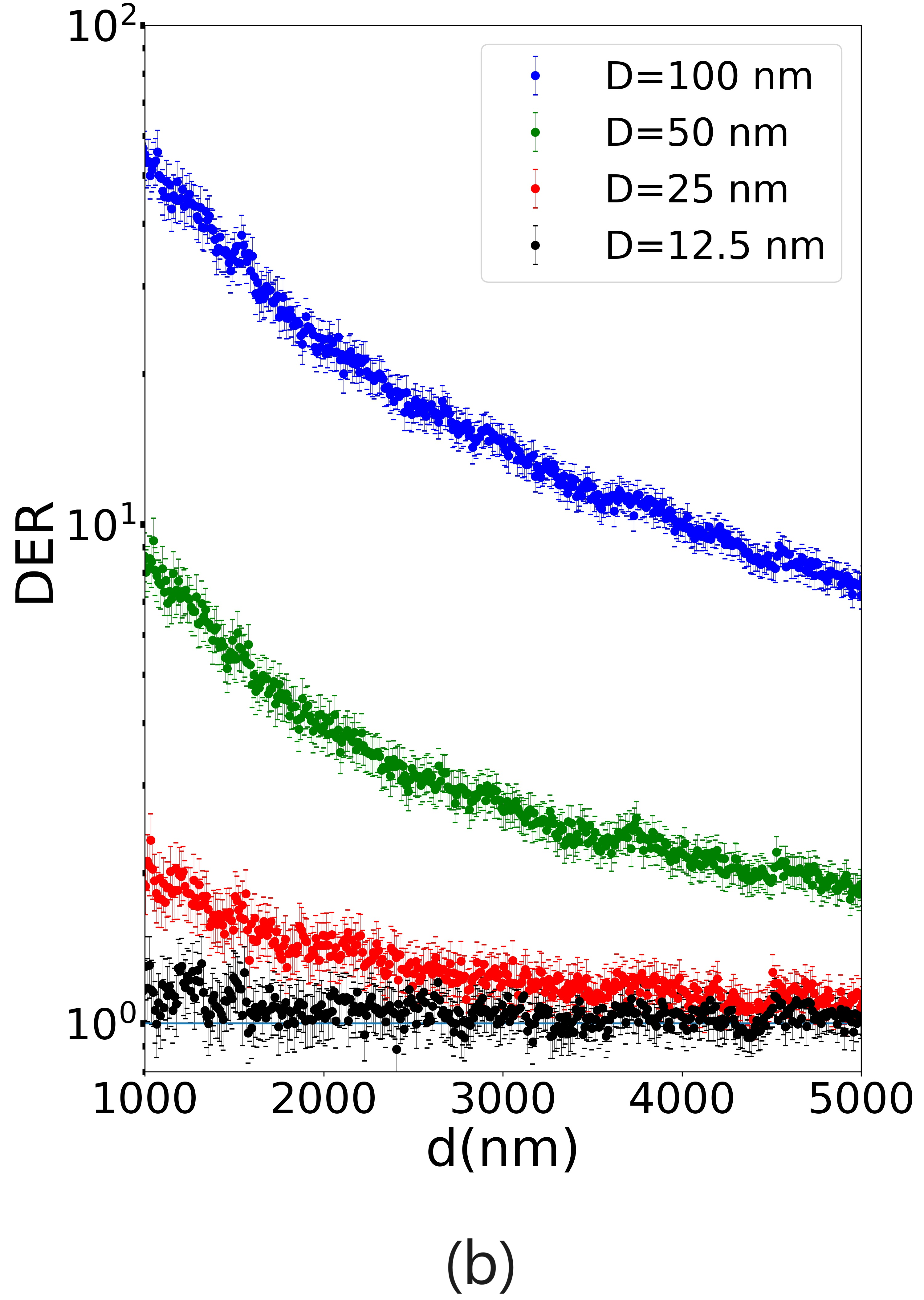}
\end{minipage}%
\hfill
\begin{minipage}[t]{0.3\textwidth}
  \includegraphics[width=1.\linewidth, height=1.3\linewidth]{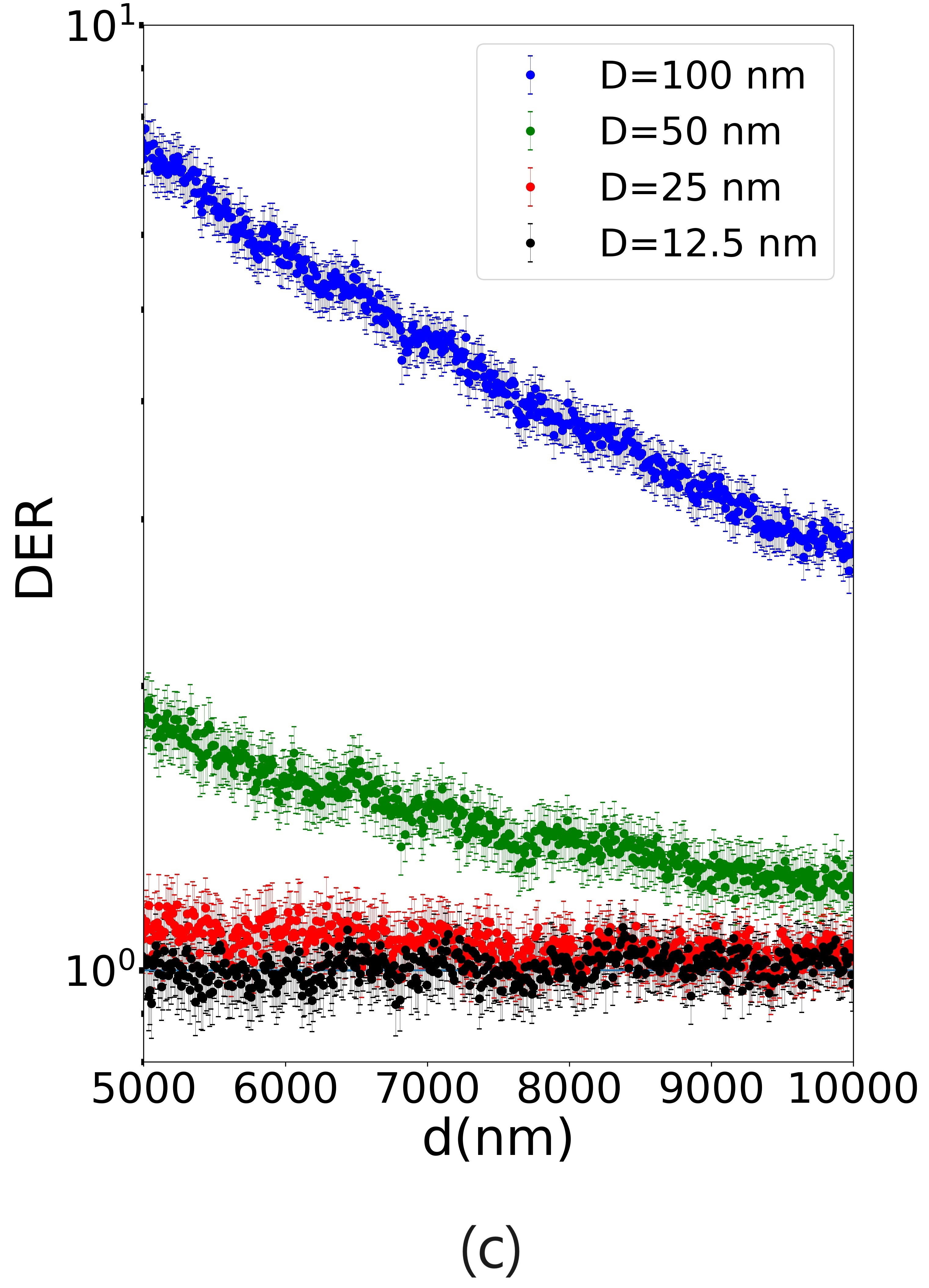}
\end{minipage}%

\caption{DER in breast medium; a) up to 1 $\unit{\micro m}$; b) from 1 $\unit{\micro m}$ to 5 $\unit{\micro m}$; c) from 5 $\unit{\micro m}$ to 10 $\unit{\micro m}$ for the 50 kVp spectrum.}
\label{fig:DER}
\end{figure*}

Results in breast medium for  AuNPs show practically no changes to their water medium counterparts. These are shown in Figure   \ref{fig:DER}. From figure \ref{fig:DER}.a) it's clear that AuNPs enhance the dose in their close vicinity, with a sharp DER decrease up to 200 nm. Larger diameters have higher DER, with energy deposition increasing 300 times at the first nanoshell. Between 200 and 700 nanometers, DER remains steady, with dose enhancement decreasing more slowly. 

In Figure \ref{fig:DER}.b), the two largest nanoparticles exhibit noticeable dose enhancement, with a fivefold DER increase when doubling the diameter, contrary to the smaller AuNPs which display minimal dose enhancement. Figure \ref{fig:DER}.c) shows that the largest AuNP affects energy deposition beyond 5 $\mu$m.  At 10 $\mu$m, it triples the dose, exceeding that of a 25 nm size at a half distance. The impact of the two smallest AuNPs is identical beyond 7 $\mu$m.

The lack of charged particle equilibrium (CPE) in DER calculations is problematic in the determination of the dose, and this issue has been brought up by several authors \cite{Rabus1, Rabus2}, who proposed several methods to avoid losing CPE conditions. In this particular work, no correction was implemented. 

\subsubsection{Surrounding Halo}

As explained in the methodology, the second PSF was used to analyse the physical processes just around the AuNP. PENELOPE PSFs output information on the position, direction,and energy of the scored particles, together with $"ILB_{i}", i=1-5$ arrays which convey information on:

\begin{enumerate}
    \item the generation of the particles (primary=1, secondary=2),
    \item the parent particle(electron=1, photon=2, positron=3),
    \item the interaction,
    \item atomic relaxation information.
\end{enumerate}

 \vspace{2mm}
 
\begin{figure}[h!]
    \centering
    \includegraphics[width=0.52\linewidth]{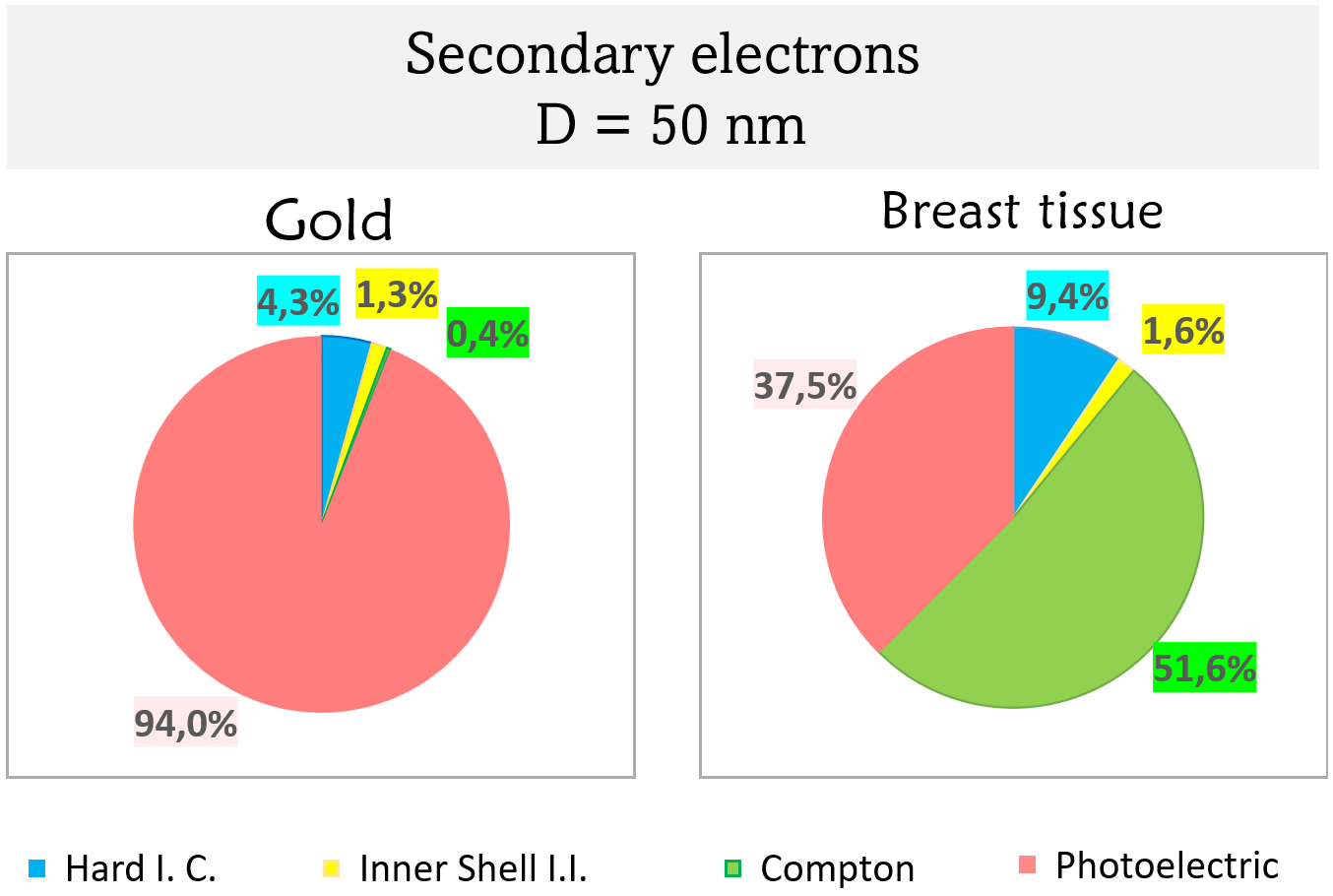} \captionof{figure}{Origin of secondary electrons.}
    \label{fig:circular}
 \end{figure}

 Figure \ref{fig:circular}  reveals the stark contrast in the interaction mechanisms in the presence or absence of an AuNP. In the absence of an AuNP,  the primary source of secondary electrons is, as expected, Compton scattering. This changes drastically when 
 in the presence of the 50 nm AuNP, in which the photoelectric absorption rises sharply and becomes the dominant mechanism for secondary electron production. This behaviour repeats for all sizes.
 
As can be seen from table \ref{tab:tissue_secondary_particles}, in breast tissue, the halo scores only  4 to 220 secondaries. Almost all are Compton, with some photo-electrons. Only in the 100 nm case can a single fluorescence event be seen. 

In stark contrast, \ref{tab:secondary_particles} shows that replacing the tissue volume by an AuNP increases the secondary yield by $\mathrm{10^{3}\text{–}10^{4}}$. This surge is driven by L-shell photo-absorption in gold (the K-shell is inactive at $<E> \approx$ 30 keV)\cite{xraydbAu}.

For instance, for the \SI{100}{nm} AuNP, ~238,000 photon parent photoelectrons are generated, representing 80\,\% of all secondaries. The ensuing Auger cascades add a comparable number of
low-energy electrons, plus \(~\sim4\times10^{4}\)  L-fluorescence photons.

\begin{figure*} [h!]
    \centering
    \includegraphics[width=0.75\textwidth]{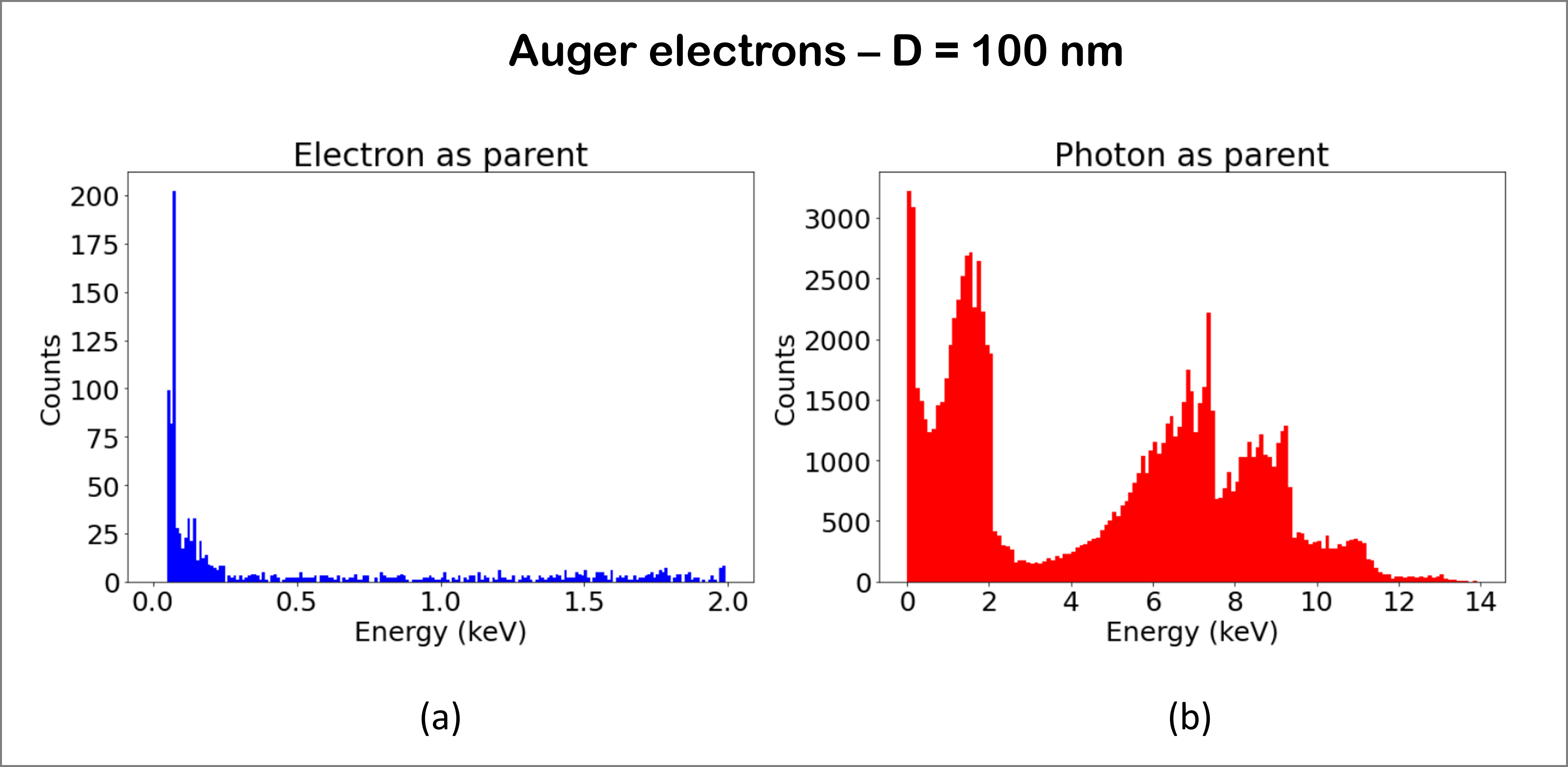}
    \caption{Auger electrons detected in the vicinity of a 100 nm diameter AuNP.}
    \label{fig:auger_el}
\end{figure*}

\vspace{1mm}
\noindent Figures \ref{fig:auger_el}.a) and b) show the production of Auger electrons in the presence of the 100 nm AuNP, revealing that Auger electrons with photon parents carry higher energies than those from electron parents. This is because the photoelectric effect at these energies will mostly lead to the ejection of an L-shell electron. In contrast, an electron parent (from Compton scattering
or hard inelastic collisions) tends to produce more shallow vacancies
or even directly ionize outer shells.

\vspace{2mm}

PENELOPE \cite{penelope} uses electron binding energies from  LLNL Evaluated Atomic Data Library (EADL)\cite{Perkins1991EADL}, and uses them to compute the Auger electron transition energies, using the first order (single-hole) approximation:

\begin{equation}
    E_e = U'_{S_0}-U'_{S_1}-U'_{S_2},
\end{equation}

where $U'_{S_0}$ is energy of the shell where the vacancy originates, $U'_{S_1}$
is the energy of the subshell that donates the electron that fills that vacancy, and $U'_{S_2}$ is the energy of the shell from which the Auger electron is ejected. This allows for an estimation of the Auger electron transition energies.

This cascading effect is made more visible by analysing PENELOPE's $\mathrm{ILB}_4$ arrays, which can be decrypted to provide a detailed picture of the  Auger emission cascades.

The decryption of the values follows the equation:  
\begin{equation}
\mathrm{ILB}_4
      = Z\,10^{6}
      + IS_1\,10^{4}
      + IS_2\,10^{2}
      + IS_3 ,
\label{eq:ILB4decode}
\end{equation}
where \(Z\) is the atomic number of the emitter
(\(Z=79\) for gold) and \(IS_{1,2,3}=1\ldots 30\) are the numerical shell labels of  (\(1=K,\;2=L_1,\;\dots\))\cite{penelope}.
Entries with \(IS_3=0\) are two-level radiative transitions (X-rays) and
were discarded from this analysis; the remaining three-level codes represent Auger
electrons.

\noindent

Table \ref{tab:au_auger} shows the top Auger transitions by ascending energy for the considered four AuNP diameters. This table makes it clear that the photoelectric driven Auger events will likely initiate with an L vacancy and de-excitation mostly via LMM → MNN → NOO cascades. 

Because the second phase-space-file registers only the electrons that \emph{exit} the
AuNP, the  LMM → MNN → NOO cascades will be affected by a competition between production and self-absorption in gold.

\begin{table}[h!]
  \centering
\begin{tabular}{cccccc}
    \toprule
    Auger & $\sim E_k$  & d = 12.5 nm & d = 25 nm & d = 50 nm & d = 100 nm \\ 
    transition & (keV) & (\%) & (\%) & (\%) & (\%) \\
    \midrule
    $\mathrm{N}_7\mathrm{O}_5\mathrm{O}_5$ & 0.07  & 6.1  &  --   &  --  &  --  \\
    $\mathrm{M}_5\mathrm{N}_3\mathrm{N}_5$ & 1.33  & --   &  7.2  &  --  &  --  \\
    $\mathrm{M}_5\mathrm{N}_4\mathrm{N}_5$ & 1.52  & 9.8  & 10.0  & 5.2  &  --  \\
    $\mathrm{M}_5\mathrm{N}_5\mathrm{N}_5$ & 1.54  & --   &  9.1  & 5.6  &  --  \\
    $\mathrm{M}_4\mathrm{N}_4\mathrm{N}_5$ & 1.61  & 6.1  &  8.4  &  --  &  --  \\
    $\mathrm{M}_4\mathrm{N}_6\mathrm{N}_7$ & 2.12  & 4.9  &  --   &  --  &  --  \\
    $\mathrm{M}_5\mathrm{N}_5\mathrm{N}_6$ & 1.78  & 6.1  &  8.6  &  --  &  --  \\
    $\mathrm{M}_5\mathrm{N}_5\mathrm{N}_7$ & 1.79  & --   &  --   & 7.5  & 5.4  \\
    $\mathrm{M}_5\mathrm{N}_6\mathrm{N}_7$ & 2.04  & 25.6 & 20.7  & 12.7 &  --  \\
    $\mathrm{M}_5\mathrm{N}_7\mathrm{N}_7$ & 2.04  & 12.2 & 16.9  & 8.0  & 6.7  \\
    $\mathrm{M}_4\mathrm{N}_6\mathrm{N}_6$ & 2.12  & 7.3  &  --   &  --  &  --  \\
    $\mathrm{M}_4\mathrm{N}_6\mathrm{N}_7$ & 2.12  & --   &  --   & 9.2  & 7.4  \\
    $\mathrm{L}_3\mathrm{M}_2\mathrm{M}_3$ & 6.03  & --   &  --   &  --  & 5.3  \\
    $\mathrm{L}_3\mathrm{M}_3\mathrm{M}_3$ & 6.43  & 4.9  &  --   & 5.5  & 7.6  \\
    $\mathrm{L}_3\mathrm{M}_3\mathrm{M}_4$ & 6.89  & --   &  --   & 6.2  & 8.1  \\
    $\mathrm{L}_3\mathrm{M}_3\mathrm{M}_5$ & 6.97  & --   &  --   & 8.2  & 11.0 \\
    $\mathrm{L}_3\mathrm{M}_4\mathrm{M}_5$ & 7.42  & --   & 12.1 & 15.4 & 20.7 \\
    $\mathrm{L}_3\mathrm{M}_5\mathrm{M}_5$ & 7.51  & 7.3  &  7.2  & 10.1 & 14.0 \\
    $\mathrm{L}_2\mathrm{M}_4\mathrm{M}_5$ & 9.24  & 9.8  &  --   & 6.4  & 8.9  \\
    $\mathrm{L}_3\mathrm{M}_5\mathrm{N}_5$ & 9.38  & --   &  --   &  --  & 5.0  \\
    \bottomrule
\end{tabular}

  \caption{Top Auger transitions (in percentage of total) in the AuNPs sorted by ascending energy order for the different diameters considered.}
  \label{tab:au_auger}
\end{table}

For the \SI{12.5}{nm} diameter AuNP, the shorter escape path allows the scoring of at least one of the end-of-line  N\(_7\)O\(_5\)O\(_5\) cascade transitions, the only size where these make it to the top 12, but only at 6.1 \%. The mid-cascade MNN transitions dominate, with the M\(_5\)N\(_6\)N\(_7\) transition standing out at 25.6 \%.

For the \SI{25}{nm} AuNP, some more lower-energy mid-cascade MNN Augers still escape,  the  leading line being M\(_5\)N\(_6\)N\(_7\) (20.7 \%), followed by M\(_5\)N\(_7\)N\(_7\) (16.7\%), and M\(_5\)N\(_4\)N\(_5\) (10.0 \%). Still, there's a first generation L\(_3\)M\(_4\)M\(_5\)  line at 12.1 \%.

In the middle-sized \SI{50}{nm} AuNP, electrons from both first and second-generation LMM and MNN transitions dominate; most energies in the spectrum  have similar intensities, except for L\(_3\)M\(_4\)M\(_5\) with 15.4 \%, and M\(_5\)N\(_6\)N\(_7\) at 12.7 \%.

Finally, for the larger \SI{100}{nm} AuNP, there's a clear hardening of the spectra, as only electrons whose CSDA range in Au exceeds \(\sim\SI{50}{nm}\) can escape; consequently, the spectrum is dominated by higher-energy first-generation LMM lines: L\(_3\)M\(_4\)M\(_5\) (20.7 \%), L\(_3\)M\(_5\)M\(_5\) (14.0 \%), L\(_3\)M\(_3\)M\(_5\) (11.0 \%). The most intense mid-cascade line, M\(_4\)N\(_6\)N\(_7\), contributes only 7.4 \%.

\begin{figure*} [h!]
    \centering
    \includegraphics[width=0.75\textwidth]{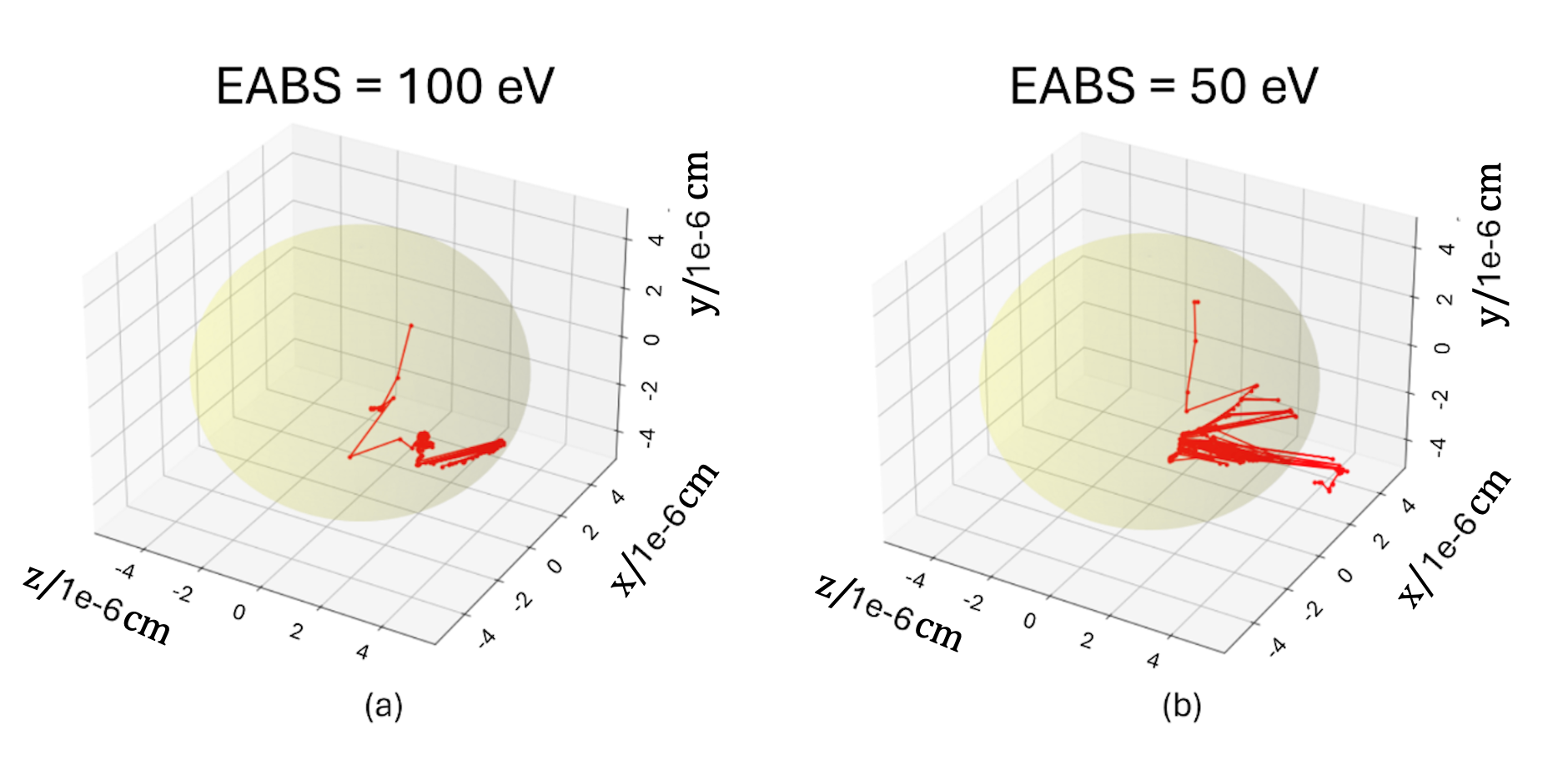}
    \caption{Track structure of two auger electrons produced in an AuNP when irradiated by a 50 kVp source, using an absorption energy of 100 eV (a) and 50 eV (b).}
    \label{fig:track}
\end{figure*}

For illustration purposes, Figure \ref{fig:track} shows the Auger track structures of two Auger electrons produced at the surface of a 100-nm AuNP. These track structures were obtained by using the Tally spatial distribution considering 180 voxels along each axis.

\subsubsection{LEM model results - survival curves}

\label{sec:LEMresults}

\begin{figure}[h!]
  \centering
  \includegraphics[width=0.8\linewidth]{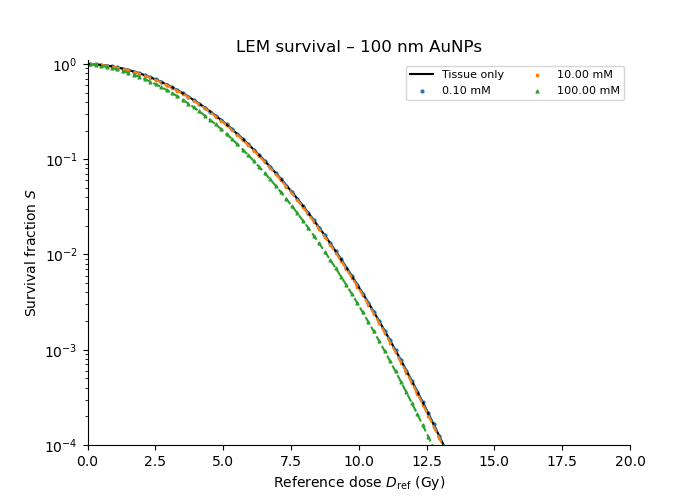}
  \captionof{figure}{Survival curve for the 100 nm AuNP for three concentrations 0.1, 10 and 100 mM}
  \label{fig:survBreast}
\end{figure}

Figure \ref{fig:survBreast} shows the Local–Effect–Model (LEM)
survival curve obtained for a breast nucleus in the presence of 100 nm AuNPs, where the sensitization is more visible.  

As explained in the methodology, the mixed term
\(2\beta\,D\,d'_i\) in the LEM expansion is discarded.
This leads to a lethal‐lesion density scaling
linearly with the prescribed dose, producing a net
shift in the linear coefficient $\alpha$ while leaving the quadratic term $\beta$
unchanged \cite{Lechtman2017}.

Table \ref{fig:ser} lists the effective
\(\alpha'\) values extracted from the log–log fits.
As predicted, the quadratic coefficient remains at the baseline value
\(\beta=0.052\;\mathrm{Gy^{-2}}\) for all curves (fit residuals \(R^{2}=1\)) and isn't shown in the table.

With the two clinically plausible AuNP concentrations (0.1–10 mM) the used LEM predicts an 
$\alpha$‐dominated radiosensitisation up to SER$_2 \approx1.1$.

At 10 mM the largest particles (100 nm) increase $\alpha$ by
$\approx\!24\%$, leading to a 4 \% drop in the survival fraction at 2 Gy ($\mathrm{SF}_2$), or a sensitisation enhancement ratio  at 2 Gy ($\mathrm{SER}_2$) of $\approx 1.04$.
This is slightly below previously published values of SER$_2=1.14–1.55$ for MV photons at a few mM intracellular gold \cite{Schmidt2022} and  Monte Carlo dose‐enhancement factors of $<$10 \% at 6–18 MV for 35–90 mM gold\cite{Mesbahi2013}.

The much softer 50 kVp beam does enhance $\alpha$,
but the effect is modest. With this model, even with an intentionally high, but clinically implausible concentration of 100 mM , where an $\mathrm{SF}_2 \approx 0.55$ and an $\mathrm{SER}_2 \approx 1.4$ were obtained, but still well below the enhancements reported for keV beams in endothelial cells\,\cite{Brown2017}. It should be noted, though, that these cells have a much more pronounced linear $\alpha$ term (roughly an order of magnitude higher) and harder x-ray diagnostic beams, well above the K-edge, plausibly explaining the higher sensitization.

\subsubsection{Distribution of AuNPs in a flattened cell}

To ensure comparability, $10^7$ histories in each simulation were used, normalizing results such that the dose in the absence of AuNPs is approximately ~1 Gy. This setup enables a direct comparison of the radiosensitisation effect for two different nanosphere sizes while maintaining a consistent gold mass.

\begin{figure}[h!]
    \centering
    \includegraphics[width=1\linewidth]{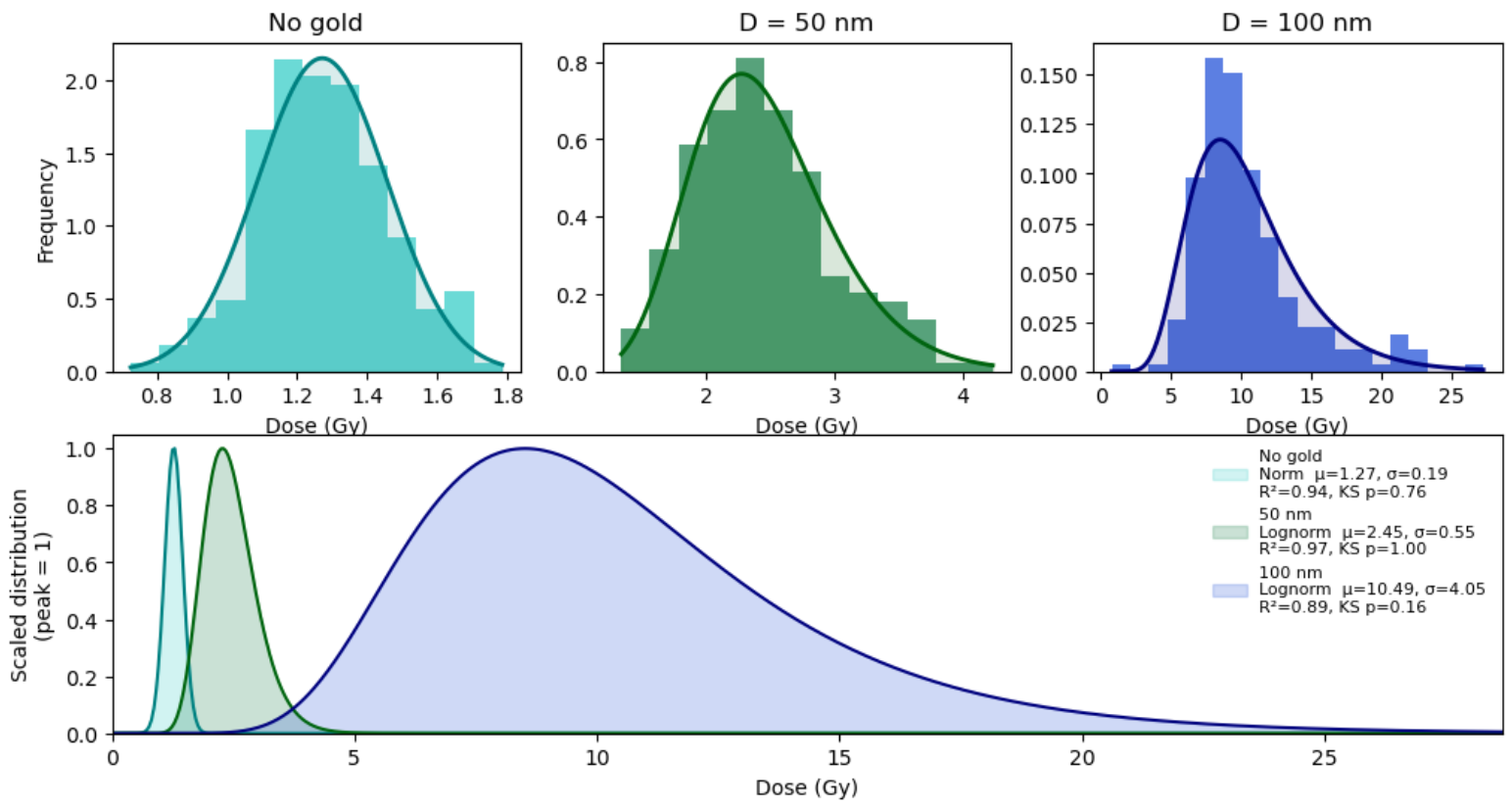} \captionof{figure}{Histograms and fitted PDFs for (a) no AuNPs and distribution of 2000 AuNPs with a diameter of (b) 50 nm  and (c) 100 nm. Below, the  distributions, all normalized to peak=1,  are shown together for comparison.}
    \label{fig:2000}
 \end{figure}

Figure~\ref{fig:2000} presents the dose distributions in the nucleus for three cases: (i) no gold nanoparticles (AuNPs), (ii) 50 nm AuNPs, and (iii) 100 nm AuNPs, with a total of 2000 nanoparticles distributed in the cytoplasm, corresponding to concentrations of 8.15 mM and 67 mM in the cytoplasm, respectively. 

The inclusion of AuNPs significantly modifies the dose distribution, clearly shifting it to higher values. 

To better understand the changes, regression curve fits, tentatively using the gamma, log-normal, Weibull, inverse-Gaussian and normal distributions were made.
Goodness-of-fit was assessed via the Akaike Information Criterion (AIC) and a Kolmogorov–Smirnov (KS) test, both available in the python scipy package.

For the water reference, all models were statistically equivalent ($\mathrm{\Delta AIC}<1$, and KS p $>$ 0.7), therefore the normal model was retained.

With 50 nm AuNP the gamma, log-normal, Weibull and inverse-Gaussian PDFs were equally supported ($\mathrm{\Delta AIC}<2$ and KS p $\geq$ 0.90), whereas the normal model was rejected ($\mathrm{\Delta AIC}=11$ and KS p $\geq$ 0.44).

For 100 nm AuNP the log-normal PDF provided the best fit ($\mathrm{\Delta AIC}>10$, and KS p $=$ 0.16), and was therefore adopted.

The adoption of the log-normal distribution is compatible with track lengths of photoelectron distributions \cite{Prieskorn:2014}. 

If we assume that the enhanced dose deposited in a scoring voxel per photon history can be described as \cite{Williamson1987}:
\begin{equation}
    D_{enh} = \frac{1}{m_{\text{vox}}}\sum_{i=1}^{N_e}
S\!\bigl(E_i\bigr)\,
\ell_i\,
g_i,
\end{equation}

in which $N_e$ is the number of Augers, $S(E_i) \equiv -\mathrm dE/\mathrm dx$ is the stopping power, $\ell_i$ the fraction of the electron’s track length that lies inside the voxel, dependent on stochastic mechanisms such as straggling and quasi-elastic interactions; and $g_i$ is the local enhancement factor, all positive and nearly independent.  As such, we can consider that their product is log-normally distributed by the geometric central-limit theorem.  

As proven in equation \ref{eq:delta_N_IV}, the mean  number of ionisations, and therefore local dose, scales with $\propto R^{3}$, and it is safe to assume it follows a Poisson distribution. 

This  means that if $ D \sim \mathrm{LogN}(\mu, \sigma^2) $, then $R^3 X \sim \mathrm{LogN}(\mu + \ln R^3, \sigma^2)$ (by the law of logarithmic multiplication). 

This provides a theoretical framework to explain the dose enhancement due to AuNP presence and the sharper shift following a log-normal with the 100 nm.

In fact, if we consider that the log-normal local dose enhancement in a voxel can also be written as:

\begin{equation}
    D_{enh} = D_0e^{\sigma Z},
    \label{D_lognorm}
\end{equation}

where $D_0$ corresponds to the baseline dose, $Z\sim \mathcal{N}(0,1)$ is normally distributed, and $\sigma$ measures how much variability the AuNP cascade adds to the dose, when $\sigma \rightarrow 0$, the equation \ref{D_lognorm} becomes

\begin{equation}
    D_{enh} \approx D_0(1+\sigma Z),
    \label{D_lognorm_approx}
\end{equation}

which is normally distributed with mean $D_0$ and we thus retrieve the normal behaviour observed in the water baseline.

This provides a physical basis for the log-normal fits observed, in particular for the 100 nm nanoparticles where \(N_e\) is large.

The mean values $\mu$ obtained for the distribution (1.27 for baseline, 2.45 for 50 nm AuNPs, and 10.49 for 100 nm AuNPs) give a DER of $\frac{2.45}{1.27}\approx 1.9$ for the 50 nm AuNPs and  $\frac{10.49}{1.27}\approx 8.3$ for the 50 nm AuNPs is consistent with the $\Delta N \propto R^3$ derived previously, as increasing the radius by double (2) increases the ionizations, therefore the DER, by (8).

It is also noteworthy that variance is growing faster than the mean, which means the heterogeneity of the dose distribution is increasing with the presence of AuNPs and their size which is compatible with the increase in Auger transitions for bigger AuNPs as shown in the PDF section.

The growth of the mean and variance   is consistent with the ionisation-volume law and the multiplicative stopping-power formalism established in equations \ref{D_lognorm}\ref{D_lognorm_approx}. The log-normal shape therefore has a clear mechanistic basis.

\paragraph{Modified LEM}

As shown in equation \ref{D_lognorm_approx} the extra energy deposited in a scoring voxel by the AuNP cascade can be approximated in first–order,
where $D_{0}\simeq 1\,$Gy is the homogeneous dose in the absence of AuNPs and
$\sigma$ encapsulates the relative width of the enhancement distribution, which will be further analysed later.

Considering again the survival fraction as given by:
\begin{equation}
    S = e^{-\alpha D_{enh} - \beta D_{enh}^2}
    \label{S_fraction_II}
\end{equation}

Expanding from equation \ref{D_lognorm_approx}, each term in the survival function becomes :
\begin{equation}
  \begin{aligned}
-\alpha D_{\text{enh}}
&= -\alpha D_{0}\bigl(1+\sigma Z\bigr)
   = -\alpha D_{0} \;-\; \alpha\sigma D_{0} Z, \\[4pt]
-\beta D_{\text{enh}}^{2}
&= -\beta D_{0}^{2}\bigl(1+2\sigma Z+\sigma^{2}Z^{2}\bigr)  \\[2pt]
&= -\beta D_{0}^{2} \;-\; 2\beta\sigma D_{0}^{2} Z
   \;-\;\beta\sigma^{2}D_{0}^{2}Z^{2}.
\end{aligned}
\end{equation}

In which we will again consider the probability of simultaneous baseline and enhanced Auger hits to be negligible, and therefore the quadratic cross-term $2\sigma Z$ can again be discarded \cite{Lechtman2017}. 
\begin{equation}
    S \approx
    e^{
      -\alpha D_{0}(1+\sigma Z)\;-\;\beta D_{0}^{2}
      \;-\;\beta\sigma^{2}D_{0}^{2}Z^{2}
    },
    \label{S_fraction_III}
\end{equation}

to note that we keep a quadratic term in $\beta$, which will be analysed later.

Averaging over the Gaussian fluctuations yields
\begin{align}
  S=\bigl\langle S_{enh} \bigr\rangle
  &= e^{-\alpha D_{0}-\beta D_{0}^{2}}\,
     \mathbb E_{Z}\!\bigl[
        e^{-\alpha\sigma D_{0}Z-\beta\sigma^{2}D_{0}^{2}Z^{2}}
     \bigr]                                                     \nonumber\\[4pt]
  &= e^{-\alpha D_{0}-\beta D_{0}^{2}}\,
     \frac{
           \exp\Bigl(
             \dfrac{\alpha^{2}\sigma^{2}D_{0}^{2}}
                   {2\bigl(1+2\beta\sigma^{2}D_{0}^{2}\bigr)}
           \Bigr)}
          {\sqrt{1+2\beta\sigma^{2}D_{0}^{2}}},
  \label{eq:Ssigma_D0}
\end{align}
where we used the Gaussian identity
$\mathbb E[e^{tZ+bZ^{2}}]=e^{\frac{\frac{t^{2}}{2(1-2b)}}{\sqrt{1-2b}}}$ for $b<\frac12$ \cite{abramowitz1965handbook}, and given that  
$b=-\beta\sigma^{2}D^{2}$ is always negative, since 
$\beta,\sigma^{2},D^{2}>0$, then  $1-2b \geq 1$, making the hard limit of validity only happen if $2\beta\sigma^{2}D^{2} \rightarrow 1$, or $\beta \approx \frac{1}{2\sigma^{2}D^{2}}$, meaning an extremely low value of $\beta$.

We are then left with:
\begin{equation}
  S_{\sigma}(D)
  \;=\;
  \frac{\displaystyle
        \exp\!\Bigl[
          -\alpha D-\beta D^{2}
          +\dfrac{\alpha^{2}\sigma^{2}D^{2}}
                 {2\bigl(1+2\beta\sigma^{2}D^{2}\bigr)}
        \Bigr]}
       {\sqrt{1+2\beta\sigma^{2}D^{2}}},
  \label{eq:Ssigma_D}
\end{equation}

Given that the mean dose can be written as 
$\langle D\rangle=D_{0}e^{\sigma^{2}/2}$, the simulated dose-enhancement ratio
${\rm DER}=\langle D\rangle/D_{0}$ can be related to  the stochastic variations $\sigma$:
\begin{equation}
  \sigma=\sqrt{2\ln({\rm DER})}.
  \label{eq:sigma_from_DER}
\end{equation}

Equation \eqref{eq:Ssigma_D} gives a closed-form
survival curve that can be applied to any scenario once DER is known.

The advantage of using this modified LEM method comes from not needing to perform complex voxel calculations. If one assumes the dose enhancement follows a log-normal distribution, as shown in Eqs.~\ref{D_lognorm}–\ref{D_lognorm_approx}, the remaining equations simply follow, allowing an easier theoretical framework to determine SER and full survival curves once $\sigma$ is obtained from Eq.~\ref{eq:sigma_from_DER}.

Finally, because the log–normal mean shifts as
\(
\mu \;\to\; \mu + \ln R^{3}
\)
when the particle radius doubles, the
dose–enhancement ratio scales as
\(
\mathrm{DER}\sim R^{3}
\).
Equation~\ref{eq:sigma_from_DER} therefore converts this purely geometric factor
into a stochastic width
\begin{equation}
    \sigma=\sqrt{2\ln R^{3}}.
\end{equation}

Feeding that $\sigma$ into Eq.~\ref{eq:Ssigma_D} yields physically informed survival curves. 

In our data ${\rm DER}=1.9$ (50\,nm) gives $\sigma=1.09$,
while ${\rm DER}=8.3$ (100\,nm) gives $\sigma=2.66$.

This new proposed LEM has some deep fundamental differences when compared with the "classical" LEM \cite{Brown2017}\cite{Lechtman2017}. First, in the classical LEM with dropped cross-term, there is only a linear shift in $\alpha$, as explained in the previous section, thus focusing on  single-hit cascades.

In this modified model, by comparison, the shift in $\alpha$ is quadratic. This acknowledges the random, Poisson-distributed nature of the local dose, with a mean (and therefore variance) scaling with $\propto R^3$. Equation \ref{S_fraction_III} shows that extra dose enters into the survival curve exponent as $-\alpha\sigma D_0 Z$, leading to an expected survival value that scales with $\alpha^2\sigma^2D_0^2$, quadratic in $\alpha$, which will in fact contribute less to the SER, compared with the classical LEM.

However, now $\beta$ acquires an effective shift, acknowledging that now a double-hit within the same cascade electron is no longer negligible, in particular for larger AuNPs. This happens as Poisson fluctuations, scaling with $R^3$, become larger. This term will now contribute to the SER in a non-negligible way.

\begin{figure}[h!]
    \centering
    \includegraphics[width=0.6\linewidth]{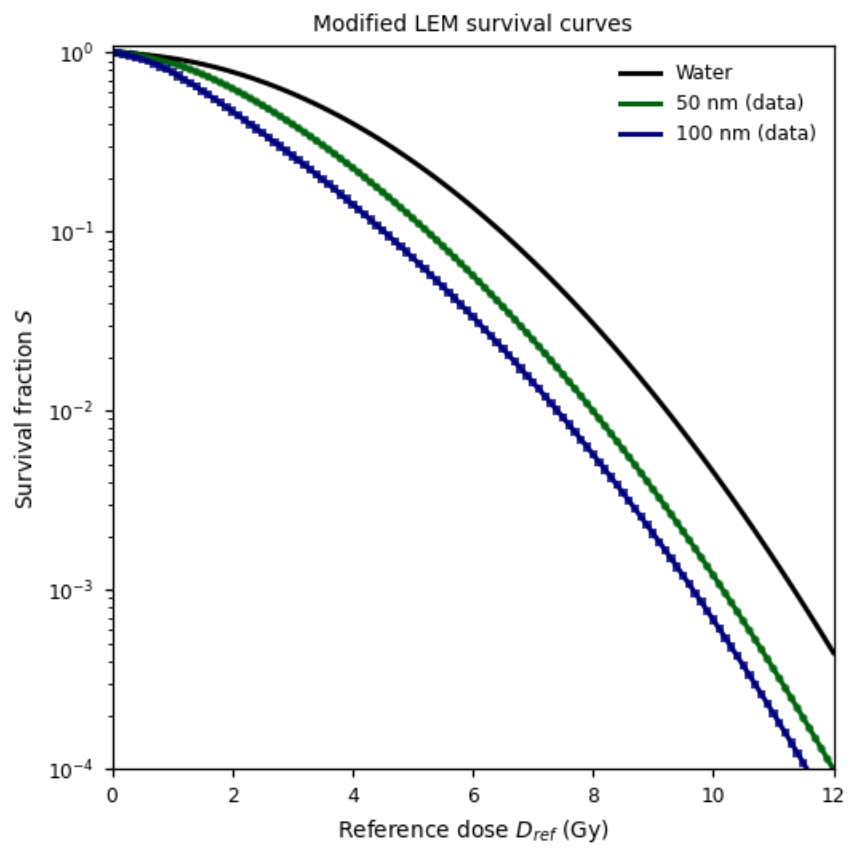} \captionof{figure}{Survival curves when modified LEM model is applied  50 nm and 100 nm data.}
    \label{fig:new_LEM}
 \end{figure}

In figure \ref{fig:new_LEM}, the modified LEM model  is applied for the previously determined DER values of 1.9 and 8.3 for 50 nm and 100 nm respectively.

The applied model gives an
$\mathrm{SER}_{2}=1.24$ for 8.15 mM concentration 50\,nm
and an $\mathrm{SER}_{2}=1.66$ for 65.2 concentration 100\,nm,
higher than the results obtained using the "classical" LEM model in the previous section.

The fitted values of $\alpha'$ and $\beta'$ are provided in table \ref{tab:fitted_params} of Supplementary Materials.

\section{Discussion}

The first objective of this work was to validate the
PENELOPE–PenEasy model.  The DER curves obtained for 50 and 100 nm AuNPs in water
(Fig.~\ref{fig:both}) agree with Li \textit{et al.}\, \cite{corrigendum}
to better than 20\,\% over the entire
range.  This agreement confirms that the
transport parameters listed in Table~\ref{tab:penelope_params} provide
an accurate description of low-energy electron cascades in the keV
domain.

Replacing water with breast medium produces DER profiles
practically indistinguishable from the water case
(Figures \ref{fig:DER} and \ref{fig:DER_water_annex}).  This was expected because the mass–density and
atomic numbers are roughly equivalent.  The four used AuNP diameters exhibit the typical observed DER pattern, which consists of an extreme near-surface enhancement (up to $\sim 3\times10^{2}$ at 12.5 nm
for 100 nm AuNPs), followed by a plateau that extends a few hundred nanometres, and
a gradual fall-off governed by the CSDA ranges of sub-20 keV electrons.
The larger spheres dominate at every distance because the number of
photo-absorptions, and therefore the released charge, scales with the
volume (\S\ref{sec:Methodology}).

Importantly, Figure \ref{fig:DER}.c shows that a single 100 nm particle
still increases the dose three-fold at 10 $\mu$m — of the same order as a
cell diameter — whereas 12–25 nm particles are indistinguishable from
the baseline beyond 7 $\mu$m .  This size dependence is a useful lever for
optimising tumour control versus healthy-tissue sparing.

The two–stage PSF analysis confirms that the physical process behind
the macroscopic DER is the large L-shell photoelectric cross-section of
gold at $E_{\mathrm{mean}}\approx31$ keV.  In the halo immediately
outside the nanoparticle, $80$–$95$\,\% of all secondaries originate
from photoelectric events (Table~\ref{tab:secondary_particles}),
whereas Compton electrons dominate in pure tissue
(Table~\ref{tab:tissue_secondary_particles}). The  vacancy set-off in the L shell provoked by the photoelectric effect leads to LMM–MNN–NOO Auger cascades, which are visible in the PSFs of the AuNPs, competing systematically with self-absorption.
(Table~\ref{tab:au_auger}). For the 12.5 nm AuNP, the full cascade is visible, but 
as self-absorption increases with size, it progressively suppresses mid- and end-cascade
lines. At 100 nm only first-generation LMM electrons
($\sim7.5$–9.5 keV) can be seen. 

In subsection 2.2.4 it is shown that the number of ionizations will be directly proportional to volume, regardless of shape. Interestingly, Taheri et al \cite{Taheri2024} made a recent study where they analyzed the influence of different nano-rod shapes in dose enhancement ratios. Their findings, in which no significant difference between  nanorods or spherical shapes can be found, further validates this point.

Applying the LEM to the nucleus-averaged dose distribution reveals an
 increase in the linear component $\alpha$, while the
quadratic term $\beta$ remains at the baseline value
(Table~\ref{fig:ser}).  This comes from neglecting 
the cross term $2\beta D d'_{i}$, making the dose enhancement linear with D.  

Using clinically realistic concentrations (0.1–10 mM) the model predicts $\mathrm{SER}_{2}$ values between 1.02 and 1.10, which is a modest yet measurable
radiosensitisation.  

The intentionally high yet experimentally unfeasible 100 mM case illustrates the
extreme limit, as $\alpha$ rises by a factor of 3.4, and the survival fraction
falls to $\mathrm{SF}_{2}\approx0.55$, with
$\mathrm{SER}_{2}\approx1.4$.  However, even in this extreme scenario, the SER is well below the
values reported for endothelial cells irradiated with
$\geq$100 keV spectra \cite{Brown2017}, where the beam energy sits above
the K-edge.

The present results therefore reinforce the view that low energy diagnostic x-ray activated AuNPs produce mainly $\alpha$ sensitisation, leading to less pronounced effects in survival fractions, scalable only by D. 

The clinical benefit thus comes from finding the proper balance between intracellular AuNP concentrations and biological limits imposed by gold uptake, excretion, and toxicity to healthy tissues.

When the classical Local Effect Model (LEM) is applied to the nucleus-averaged dose, it points to a modest but measurable radiosensitisation.  For AuNP concentrations between 0.1 and 10 mM the predicted survival-enhancement ratio at 2 Gy never exceeds about ten per cent, a result that matches the small rise in the linear component of the survival curve while the curvature remains essentially unchanged.  The trend confirms that, under low-energy irradiation, dose amplification is driven mainly by single, highly localised interactions rather than by multiple, simultaneous lesions.

Introducing the modified LEM (mLEM) allows for a more mechanistic treatment of the local enhancement, by assuming a log-normal distribution local dose. Even with the limited in-silico data available so far, the mLEM predicts an $\mathrm{SER_2}$ of roughly 1.3 for 8.15 mM of 50 nm particles and about 1.7 for 65.2 mM 100 nm.

The new mLEM has the advantage that only the DER is needed, after that, survival curves can be determined. It will be interesting to see if the model can be further replicated with different AuNPs sizes and cellular shapes, both from in-vitro and in-silico scenarios.

\section{Conclusion}

Low-energy (50 kVp) x-ray activated AuNPs increase dose deposition in their vicinity mostly through Auger LMM–MNN–NOO cascades derived from a heightened photoelectric cross-section of Gold and high Auger emission yields at these energies. 

The Dose Enhancement Ratio (DER) was determined  in water (for validation) and in breast medium. The obtained values in the breast medium show no significant changes to the results in water.

The Local Effect Model (LEM) used in this work with 50 nm AuNPs yielded survival curves that become steeper with AuNP concentration. The spatially averaged dose allowed for the determination of survival curves, showing only modest SER, so even a DER of 1.9 (50 nm AuNP) or 8.3 (100 nm AuNP) translates into at most a 10 \% increase in cell kill ($\mathrm{SER_2}
  \leq 1.10$). 100 nm particles, with a concentration of 10 mM, make the LEM's linear $\alpha$ rise by 24 \%, and survival at 2 Gy falls by only 4 \% , giving an $\mathrm{SER_2} =1.04$).

The analysis of the dose distributions inside the flattened cell allowed for the development of a modified Local-Effect Model (mLEM) which works directly with the probability distributions.  By converting the measured dose-enhancement ratios (DER) into the width of the local-dose distribution, the model can make better predictions of local sensitization, providing a direct link between 
+$\sigma = 1.09$ for \SI{50}{\nano\metre} particles and $\sigma = 2.66$ for \SI{100}{\nano\metre} particles and the AuNP radius.  

With these $\sigma$ values, the model predicts a sizeable radiosensitisation of $\mathrm{50}$ nm AuNP at \SI{8.15}{\milli\Molar} of $\mathrm{SER}_{2} = 1.30$, whereas for the $\mathrm{50}$ nm AuNP at \SI{65.2}{\milli\Molar} of $\mathrm{SER}_{2} = 1.70$.

Further work in this model is still necessary, but the results presented are promising and could perhaps set a new path for the development of more physically informed models.

\section*{Funding}
Pedro Teles thanks FCT for funding CEECINST/00133/2018/CP1510/CT0001

\section*{Acknowledgments}

Pedro Teles thanks Weibo Li for making the raw PENELOPE data from their intercomparison available.

\clearpage
\section*{Supplementary Material}
\appendix

\section{X-ray spectra}

\begin{figure}[htbp]
  \centering
  \begin{subfigure}[t]{0.48\textwidth}
    \centering
    \includegraphics[width=\textwidth]{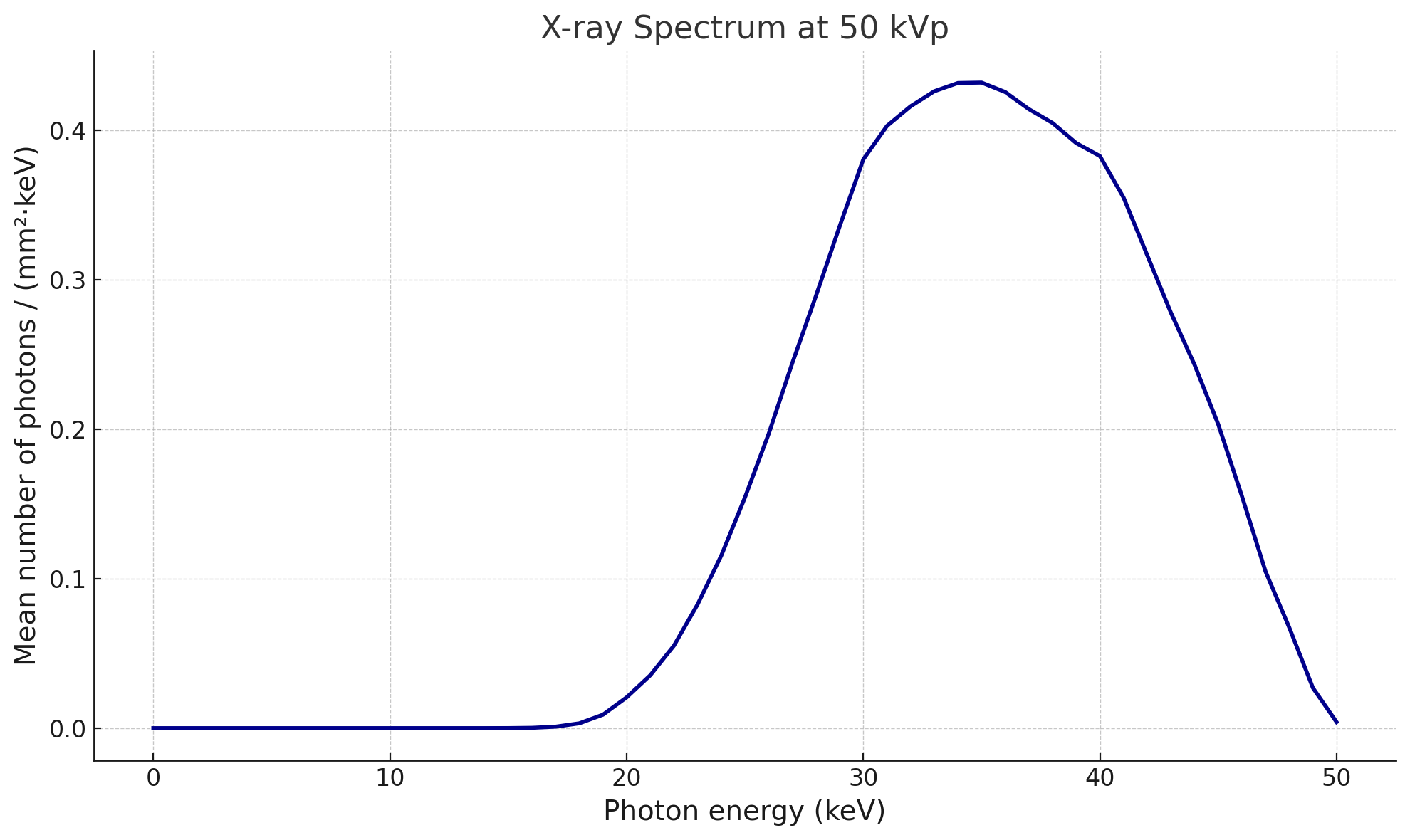}
    \caption{X-ray spectrum at 50 kVp.}
    \label{fig:spectrum50}
  \end{subfigure}
  \hfill
  \begin{subfigure}[t]{0.48\textwidth}
    \centering
    \includegraphics[width=\textwidth]{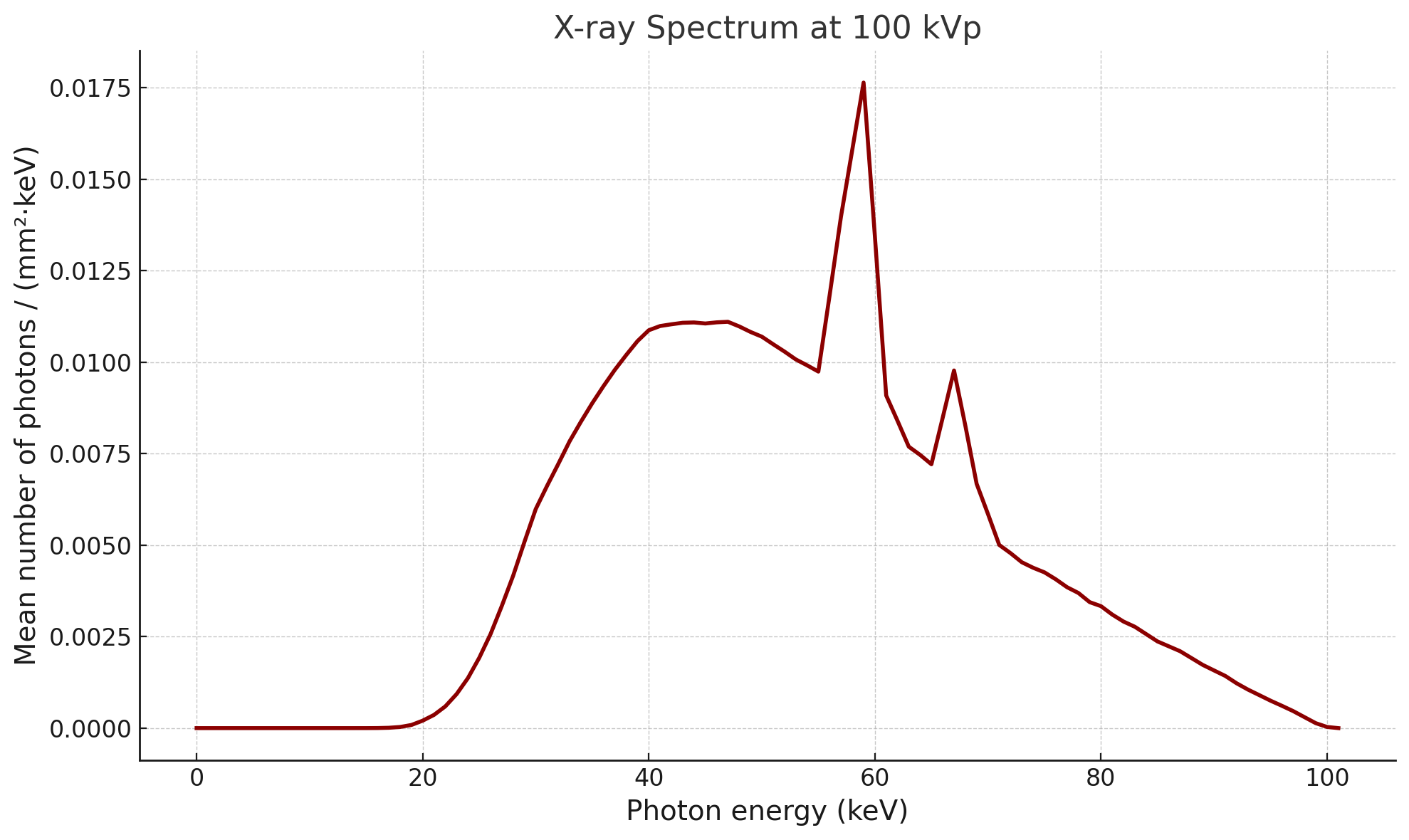}
    \caption{X-ray spectrum at 100 kVp.}
    \label{fig:spectrum100}
  \end{subfigure}
  \caption{The used X-ray spectra at 50 kVp and 100 kVp in this work.}
  \label{fig:xray_comparison}
\end{figure}

\section{PENELOPE Simulation Parameters}
\renewcommand{\thetable}{B.1} 
\begin{table}[h]
    \centering
    \scriptsize 
    \setlength{\tabcolsep}{4pt} 
    \renewcommand{\arraystretch}{1.1} 
    \caption{Penelope simulation parameters for different materials.}
    \begin{tabular}{lcccccccc}
        \hline
        Material & $\mathrm{E_{ABS}}$(e\textsuperscript{-}) (eV) & $\mathrm{E_{ABS}}$(ph) (eV) & $\mathrm{E_{ABS}}$(e\textsuperscript{+}) (eV) & C1 & C2 & WCC (eV) & WCR (eV) & $\mathrm{DS_{MAX}}$ (cm) \\
        \hline
        Gold (Au) & 50 & 50 & 50 & 0.0001 & 0.0001 & 0.5 & 0.05 & $1.0 \times 10^{-5}$ \\
        Water & 50 & 50 & 50 & 0.0001 & 0.0001 & 0.5 & 0.05 & $1.0 \times 10^{-5}$ \\
        Breast medium & 50 & 50 & 50 & 0.0001 & 0.0001 & 0.5 & 0.05 & $1.0 \times 10^{-5}$ \\
        \hline
    \end{tabular}
    \label{tab:penelope_params}
\end{table}

\clearpage

\section{DER obtained for 50 and 100 nm AuNPs in water for both spectra}

\begin{figure}[h!]
    \centering
    \includegraphics[width=0.7\linewidth]{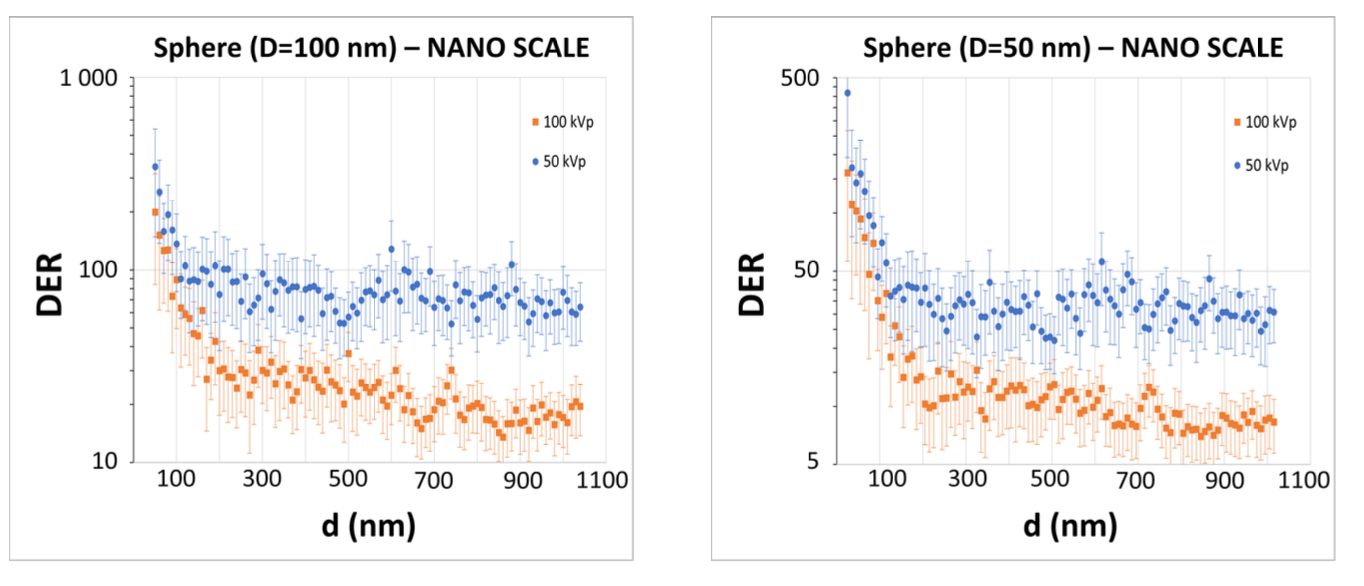} \captionof{figure}{DER obtained for (a) the 50 nm AuNP and (b) the 100 AuNP for both used spectra.}
    \label{fig:DER_water_annex}
 \end{figure}

\section{Secondary particle scoring in the "surrounding halo"}
\renewcommand{\thetable}{D.1} 
\begin{table}[h]
    \centering
    \scriptsize
    \setlength{\tabcolsep}{4pt}
    \renewcommand{\arraystretch}{1.2}
    \caption{Secondary particle generation in breast medium for different diameters in the concentric shell surrounding the particle (in this case in the absence of a particle) }
    \begin{tabular}{lcccc}
        \hline
        \textbf{TISSUE} & \textbf{D = 100 nm} & \textbf{D = 50 nm} & \textbf{D = 25 nm} & \textbf{D = 12.5 nm} \\
        \hline
        Total of primary particles & 92749108 & 23190569 & 5798902 & 1449218 \\
        Mean energy of secondary electrons (keV) & $\sim 10.7$ & $\sim 11.8$ & $\sim 8$ & $\sim 13$ \\
        Mean energy of secondary photons (keV) & $\sim 0.5$ & - & - & - \\
        \hline
        \textbf{Electrons as parent} & & & & \\
        \quad Sec. Electron: Hard I.C. & 5 & 6 & - & - \\
        \quad Sec. Electron: Inner-shell I.I. & 1 & 1 & - & - \\
        \hline
        \textbf{Photon as parent} & & & & \\
        \quad Sec. Electron: Compton & 135 & 33 & 11 & 3 \\
        \quad Sec. Electron: Photoelectric & 78 & 24 & 4 & 1 \\
        \quad Sec. Photons: Photoelectric & 1 & - & - & - \\
        \hline
        Total of secondary particles & 220 & 64 & 15 & 4 \\
        \hline
    \end{tabular}
    \label{tab:tissue_secondary_particles}
\end{table}

\renewcommand{\thetable}{D.2} 
\begin{table}[h]
    \centering
    \scriptsize
    \setlength{\tabcolsep}{4pt}
    \renewcommand{\arraystretch}{1.2}
    \caption{Secondary particle generation for AuNPs of different diameters in the concentric shell surrounding the particle.}
    \begin{tabular}{lcccc}
        \hline
        \textbf{GOLD} & \textbf{D = 100 nm} & \textbf{D = 50 nm} & \textbf{D = 25 nm} & \textbf{D = 12.5 nm} \\
        \hline
        Total of primary particles & 92615594 & 23174117 & 5797321 & 1449116 \\
        Mean energy of secondary electrons (keV) & $\sim 12$ & $\sim 10.5$ & $\sim 8.9$ & $\sim 7.5$ \\
        Mean energy of secondary photons (keV) & $\sim 9.6$ & $\sim 9.5$ & $\sim 9.5$ & $\sim 9.1$ \\
        \hline
        \textbf{Electrons as parent} & & & & \\
        \quad Sec. Electron: Hard I.C. & 8612 & 1587 & 214 & 18 \\
        \quad Sec. Electron: Inner-shell I.I. & 3638 & 528 & 65 & 4 \\
        \quad Sec. Photons: Hard Brems. Em. & 187 & 13 & 2 & - \\
        \quad Sec. Photons: Inner-shell I.I. & 96 & 4 & 1 & - \\
        \hline
        \textbf{Photon as parent} & & & & \\
        \quad Sec. Electron: Compton & 222 & 55 & 16 & 3 \\
        \quad Sec. Electron: Photoelectric & 238032 & 35142 & 4441 & 333 \\
        \quad Sec. Photons: Photoelectric & 40387 & 5030 & 514 & 34 \\
        \hline
        Total of secondary particles & 291174 & 42359 & 5253 & 392 \\
        \hline
    \end{tabular}
    \label{tab:secondary_particles}
\end{table}

\clearpage

\section{Fitted linear $\alpha$ values for  survival curves}
\renewcommand{\thetable}{E.1} 

\begin{table}[h]
  \centering

  \begin{tabular}{lccc}
    \hline
    NP diameter & \multicolumn{3}{c}{$\alpha'$ (Gy$^{-1}$)}\\
    \cline{2-4}
                & 0.1 mM & 10 mM & 100 mM\\
    No AuNP & 1.900$\times10^{-2}$ & 1.900$\times10^{-2}$ & 1.900$\times10^{-2}$\\
    \hline
    12 nm  & 1.900$\times10^{-2}$ & 1.923$\times10^{-2}$ & 2.126$\times10^{-2}$\\
    25 nm  & 1.901$\times10^{-2}$ & 1.929$\times10^{-2}$ & 2.191$\times10^{-2}$\\
    50 nm  & 1.901$\times10^{-2}$ & 1.979$\times10^{-2}$ & 2.688$\times10^{-2}$\\
    100 nm & 1.905$\times10^{-2}$ & 2.351$\times10^{-2}$ & 6.406$\times10^{-2}$\\
    \hline
  \end{tabular}
  \captionof{table}{Fitted parameters ($\alpha$ and $\beta$) for 0.1, 10 and 100 mM concentrations and the different AuNP sizes considered.}
    \label{fig:ser}
\end{table}

\section{Fitted linear $\alpha$ and $\beta$ values for the elongated cell}
\renewcommand{\thetable}{F.1} 

\begin{table}[h]
  \centering
  \begin{tabular}{lccc}
    \hline
    NP diameter & $\alpha'$ (Gy$^{-1}$) & $\beta'$ (Gy$^{-2}$) & $R^2$ \\
    \hline
    Water   & 0.0190 & 0.0520 & 1.000 \\
    50 nm   & 0.1719 & 0.0500 & 1.000 \\
    100 nm  & 0.3123 & 0.0418 & 1.000 \\
    \hline
  \end{tabular}
  \captionof{table}{Fitted parameters $\alpha'$, $\beta'$, and $R^2$ for water and AuNPs of 50 and 100 nm diameter.}
  \label{tab:fitted_params}
\end{table}

\clearpage

\section{Gold Nanoparticle Concentrations in Cytoplasm}
\label{sec:AuNP_concentration_corrected}

This section presents the calculations for the concentration of 2000 gold nanoparticles (AuNPs) in the cytoplasm, excluding the nucleus, for two different particle sizes: 50 nm and 100 nm. Calculations are done with three significant figures. We calculate for 50 nm and scale for the 100 nm knowing that it scales with $2^3$ (twice the radius cubed)

\subsection{Volume Calculations}
The volume of a single spherical AuNP is given by:
\begin{equation}
    V = \frac{4}{3} \pi r^3
\end{equation}
For 50 nm AuNPs: $r = 25$ nm $= 25 \times 10^{-9}$ m 

The volume of a single 50 nm nanoparticle:
\begin{align}
    V_{50} &= \frac{4}{3} \pi (25 \times 10^{-9})^3 = 6.54 \times 10^{-23} \text{ m}^3 \\
\end{align}

For 2000 nanoparticles, the total gold volume is given by:
\begin{align}
    V_{\text{total}, 50} &= 2000 \times 6.54 \times 10^{-23} = 1.31 \times 10^{-19} \text{ m}^3, \\
\end{align}

and the cytoplasm volume (after subtracting the nucleus) is:
\begin{equation}
    V_{\text{cytoplasm}} = 1.57 \times 10^{-16} \text{ m}^3.
\end{equation}

\subsection{Mass Concentration Calculations}
The density of gold is:
\begin{equation}
    \rho_{\text{Au}} = 19.3 \times 10^3 \text{ kg/m}^3
\end{equation}

The total mass of gold in the cytoplasm:
\begin{align}
    m_{50} &= V_{\text{total}, 50} \times \rho_{\text{Au}} = (1.31 \times 10^{-19}) \times (19.3 \times 10^3) = 2.53 \times 10^{-15} \text{ kg} \\
\end{align}

Finally, convert to mg/g:
\begin{align}
    C_{\text{mass}, 50} &= \frac{m_{50}}{V_{\text{cytoplasm}}\cdot\rho_{water}} \times 10^3 = \frac{2.53 \times 10^{-15}}{1.57 \times 10^{-13}} \times 10^3 \approx 16.1 \text{ mg/g} \\
    C_{\text{mass}, 100} &= 2^3 \times C_{\text{mass}, 50} \approx 128.8 \text{ mg/g}
\end{align}

\subsection{Molar Concentration Calculations}
Gold's molar mass is:
\begin{equation}
    M_{\text{Au}} = 197 \text{ g/mol}
\end{equation}

The number of moles of gold in the cytoplasm:
\begin{align}
    n_{50} &= \frac{m_{50}}{M_{\text{Au}}} = \frac{2.53 \times 10^{-15} \text{ g}}{197} = 1.28 \times 10^{-14} \text{ mol} \\
\end{align}

Now, divide by the cytoplasm volume:
\begin{align}
    C_{\text{molar}, 50} &= \frac{n_{50}}{V_{\text{cytoplasm}}} = \frac{1.28 \times 10^{-15}}{1.57 \times 10^{-13}} = 8.15 \text{ mM} \\
    C_{\text{molar}, 100} &= 2^3 \times C_{\text{molar}, 50} = 65.2  \text{ mM}
\end{align}

\clearpage

\begingroup
\bibliographystyle{elsarticle-num}
\bibliography{refs.bib}
\endgroup


\end{document}